# DesignCon 2018

# Introduction to Non-Invasive Current Estimation (NICE)

Jonathan L. Fasig, Mayo Clinic

Christopher K. White, Mayo Clinic
white.christopher1@mayo.edu

Barry K. Gilbert, Mayo Clinic
gilbert.barry@mayo.edu, 507-284-4056

Clifton R. Haider, Mayo Clinic
haider.clifton@mayo.edu

## Abstract

It is notoriously difficult to measure instantaneous supply current to a device such as an ASIC, FPGA, or CPU without also affecting the instantaneous supply voltage and compromising the operation of the device [21]. For decades designers have relied on rough estimates of dynamic load currents that stimulate a designed Power Delivery Network (PDN). The consequences of inaccurate load-current characterization can range from excessive PDN cost and lengthened development schedules to poor performance or functional failure. This paper will introduce and describe a method to precisely determine time-domain current waveforms from a pair of measured time-domain voltage waveforms. This Non-Invasive Current Estimation (NICE) method is based on established two-port network theory along with component and board modeling techniques that have been validated through measurements on demonstrative circuits. This paper will show that the NICE method works for any transient event that can be captured on a digital oscilloscope. Limitations of the method and underlying measurements are noted where appropriate. The method is applied to a simple PDN with an arbitrary load, and the NICE-derived current waveform is verified against an independent measurement by sense resistor. With careful component and board modeling, it is possible to calculate current waveforms with a root mean square error of less than five percent compared to the reference measurement. Current transients that were previously difficult or impossible to characterize by any means can now be calculated and displayed within seconds of an oscilloscope-trigger event by using NICE. ASIC and FPGA manufacturers can now compute the start-up current for their device and publish the actual waveform, or provide a piecewise-linear SPICE model (PWL source) to facilitate design and testing of the regulator and PDN required to support their device.

## Biographies

**Jonathan L. Fasig** received a BSEE from Milwaukee School of Engineering (Milwaukee, WI). Jonathan is currently a Principal Engineer at the Mayo Clinic Special Purpose Processor Development Group where his focus has been power delivery network design, voltage regulator design, and signal integrity analysis.

**Christopher K. White** received his BSEE from Michigan Technological University (Houghton, MI) and MSEE from the University of Minnesota (Minneapolis, MN). Chris is currently a Senior Engineer at the Mayo Clinic Special Purpose Processor Development Group. His focus is on fiber optics, power integrity, and signal integrity.

**Barry K. Gilbert** received a BSEE from Purdue University (West Lafayette, IN) and a Ph.D. in physiology and biophysics from the University of Minnesota (Minneapolis, MN). He is currently Director of the Special Purpose Processor Development Group, directing research efforts in high-performance electronics and related areas.

**Clifton R. Haider** received a Bachelor of Science in Biomedical Engineering from the University of Iowa (Iowa City, IA) and a Ph.D. in Biomedical Engineering from Mayo Clinic (Rochester, MN). He is currently Deputy Director of the Special Purpose Processor Development Group, where he directs research efforts in high performance electronics and related areas.

## Introduction

As ASIC, FPGA, and CPU dissipation levels approach 300 watts and beyond, system designers are confronted with the need to understand exactly how much current these devices will require. This need raises a fundamental question: What is current and how is it measured?

Physics tells us that electric current is the net movement of charge, and current is defined in terms of coulombs of charge moving past a point in one second of time [1]. At present there is no practical[1] way to measure charge motion directly, so it is necessary to observe the effects of charge motion and calculate the current by one of the following methods:

1. Force the current to flow through a known impedance (like a resistor R) and measure the voltage-drop (V) across that impedance. Then from Ohm's Law:

$$\boldsymbol{I = \frac{V}{R}}$$

   Current measurements using a conventional voltmeter are typically made by this method.

2. Force the current to flow along a known path (like a wire) and measure the magnetic field (B) that encircles the wire. Then from Ampere's Law:

$$\boldsymbol{I = \frac{1}{\mu o} \oint B \cdot dl}$$

   Measurements using a "clamp-on current meter" or "current probe" are made by this method.

Unfortunately, these methods can be difficult or impossible to apply.

- Ohm's Law requires the engineer to invade (break) the circuit and insert the known impedance (current-sense resistor) in series with the load. The resulting voltage drop across the sense element can increase the supply-voltage variation observed at the load and affect the operation of the device. The sense resistor also dissipates power, which may be objectionable at high current levels. This approach becomes impractical when devices with multiple power/ground pins and/or multiple-capacitor decoupling networks are involved because the number of sense-elements and corresponding voltage-drops would be very large.
- Ampere's Law requires that the current-carrying conductor be isolated and inserted into the probe so the field can be measured. This approach is problematic when the conductor is a copper-foil shape embedded in a printed circuit board.

Fortunately, network theory offers an alternate way to calculate current without invading (opening) the circuit or isolating the current-carrying conductors. For the circuit shown in **Figure 1**, the relationships between port voltages and port currents are given by equation **(1)** and equation **(2)**[2]. Notice that currents I1 and I2 can be calculated from measured voltages V1 and V2 if the network admittance parameters Y11 through Y22 are known. Non-Invasive Current Estimation (NICE) exploits this relationship to calculate time-domain current waveforms from time-domain voltage measurements.

---

[1] Direct measurement of charge motion has a somewhat colorful history that dates back to the 1840's and involves the deposition rate for electroplating metal in a solution of silver nitrate by application of a constant current. A summary is contained in NIST/NBS Circular Number 60, "Electrical Units and Standards" available at https://sizes.com/library/USA/NIST/NBS_Circ_60.htm#pg18

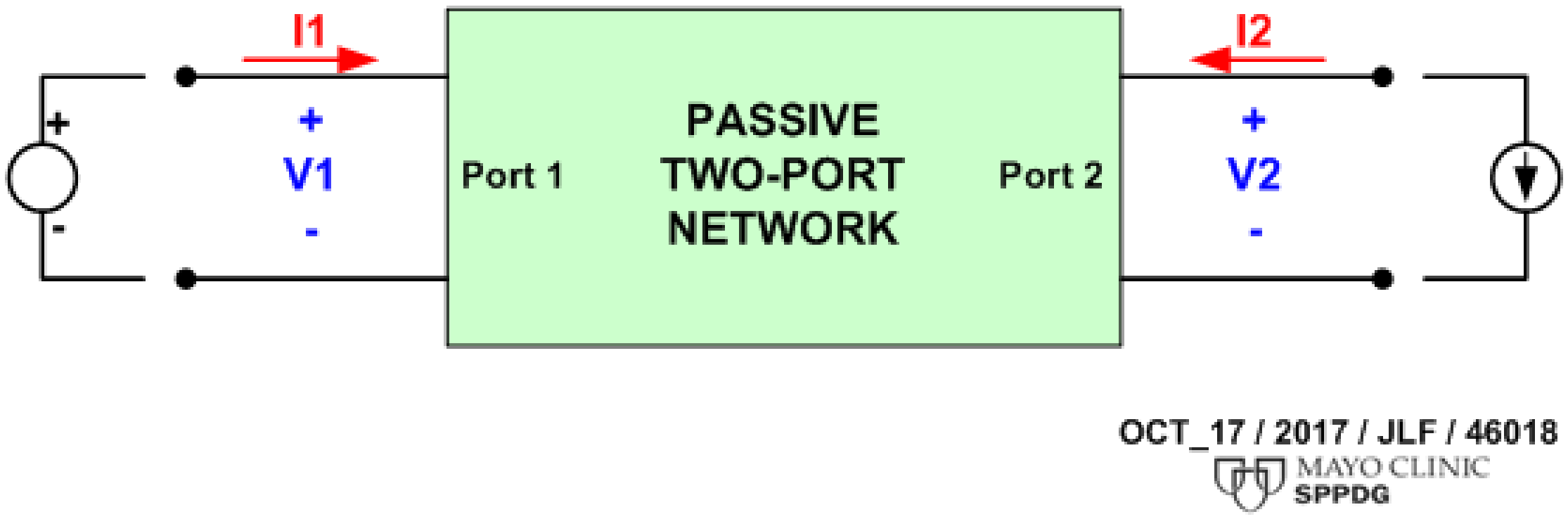

**Figure 1 – Voltage and current definitions for a two-port network**

$$I1 = Y11 * V1 + Y12 * V2 \qquad (1)$$

$$I2 = Y21 * V1 + Y22 * V2 \qquad (2)$$

While a thorough review of two-port networks is beyond the scope of this discussion, there is a special case that often recurs in power-delivery development and deserves special attention here.  That special case is the "Pi" network illustrated in **Figure 2**.

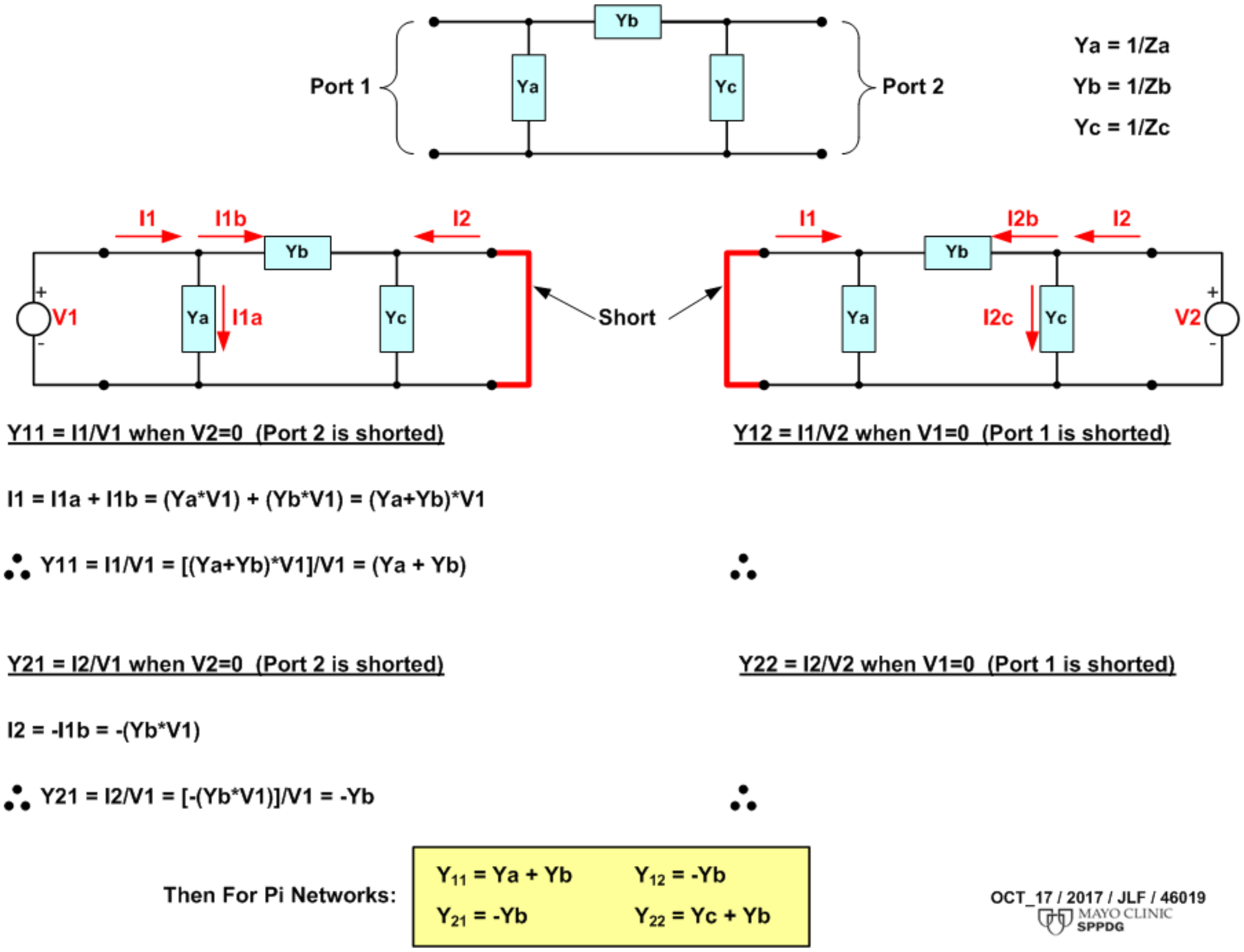

**Figure 2 - The Pi Network**

Network admittance (Y) parameters are defined by the application of a short-circuit condition at one of the network ports. This condition gives rise to the common moniker "short-circuit admittance parameters" applied to Y11, Y12, Y21, and Y22. Also notice that the admittance parameters of a Pi network can be read by inspection of the circuit diagram: the input (Y11) and output (Y22) admittances are always the sum of the shunt-admittance directly across the port plus any series admittance between the ports, while the transfer admittances (Y12 & Y21) are always minus one times the series admittance between the ports.

Conceptually, Yc can be regarded as the total admittance of the capacitors located at or near the load, while Ya can be regarded as the admittance of capacitors located away from the load (i.e., near the voltage regulator), and Yb is the admittance of the physical interconnect elements between Ya and Yc. In practice, the exact location of each capacitor and its contribution to the network are handled by extracting the network parameters from the circuit-board layout using an electromagnetics analysis tool.

The equations for Y11 through Y22 listed in **Figure 2** are valid at all frequencies, but note that each admittance parameter may be a function of frequency. Except for very simple networks with just a few components, the manual calculation of Y-parameters can become laborious, and various analysis tools and instruments are employed to facilitate the task. Fortunately, Y-parameter values can be calculated from Scattering (S) parameter data [3], and S-parameter component models are available from many manufacturers [4, 5, 6, 7]. Likewise, instruments and analysis tools that generate or manipulate S-parameters have become common, so that the creation of the Y-parameter data necessary for the NICE method amounts to either a measurement of the circuit with a Vector Network Analyzer (VNA) or collection of the relevant S-parameter models and one or more simulations using an RF-aware[2] analysis tool. In many cases, a combination of measurement and simulation may be necessary.

An example of load-current calculation from voltage measurements and network parameters is depicted in **Figure 3**. Tabulated S-parameters for a hypothetical PDN are derived by an AC frequency-sweep simulation of the circuit and subsequently converted to Y-parameters. A transient simulation using the same PDN and an arbitrary load current provide voltage waveforms for V1(t) and V2(t), and application of the Fast Fourier Transform (FFT) yields the frequency-domain signals V1(s) and V2(s). The frequency-domain load current I2(s) is calculated as shown and the Inverse Fast Fourier Transform (IFFT) is performed to produce a predicted time-domain I2(t) that is compared to the original I2 load applied during the transient simulation. The excellent correlation between the original load current and predicted load current demonstrates the feasibility of the NICE method.

[2] As used here, an "RF-aware" analysis tool is one that is capable of performing time-domain simulations using frequency-domain models such as Touchstone S-parameter files. Those tools include Keysight ADS, Synopsis Hspice, and Ansys Electronics Desktop to name a few. At the time of this writing, LTspice and Pspice do not provide this capability.

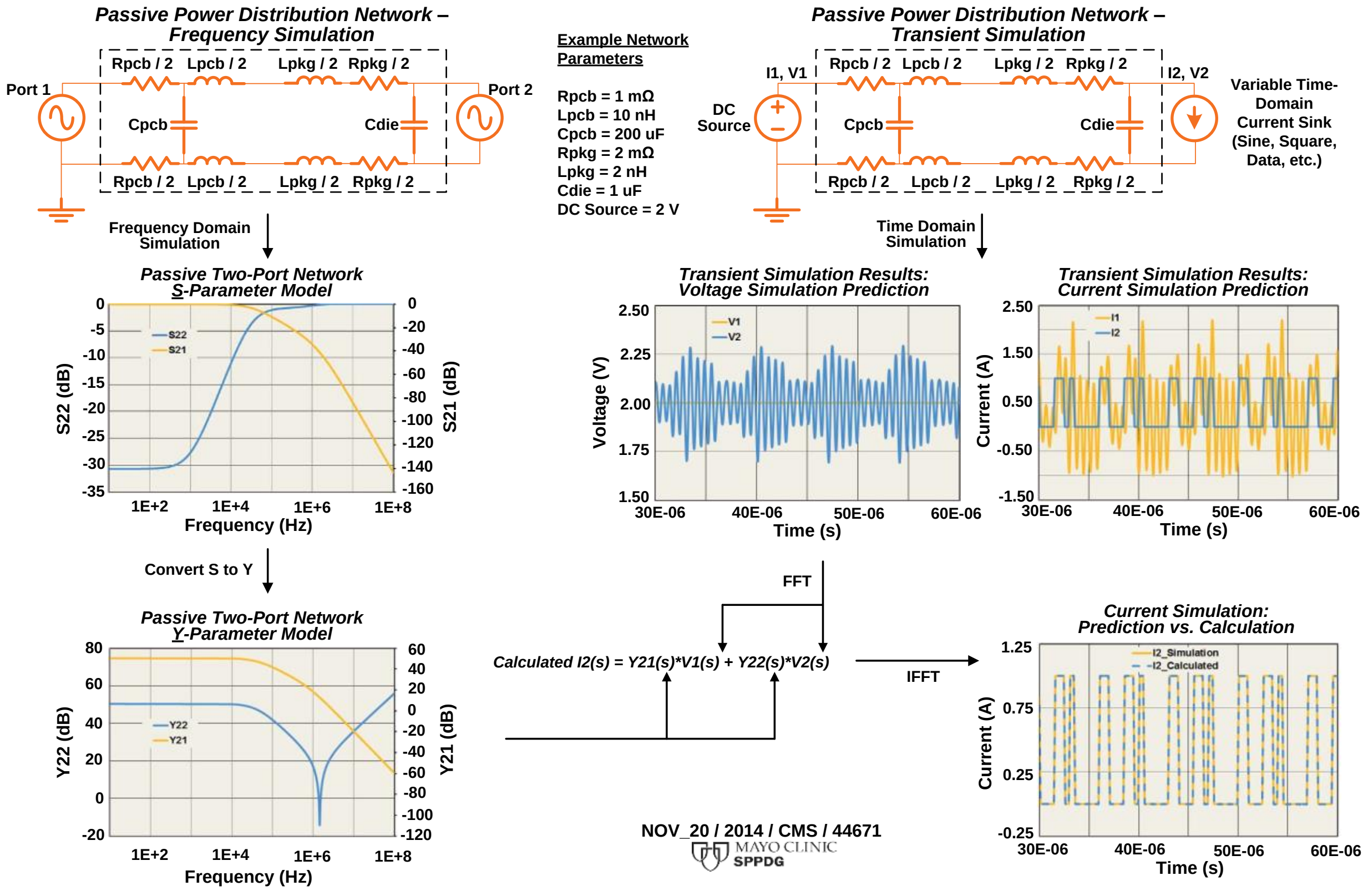

**Figure 3 - A Method to Calculate Load Current Using Two-Port Network Theory**

# Modeling

From the previous discussion, it is clear that the currents into each network port can be calculated from the voltage at each port using the admittance parameters for the network. For Power Delivery Network (PDN) designs, the network in question consists of the printed circuit board along with all of the decoupling capacitors that are connected to the power-supply net. Since the accuracy of the NICE calculation depends on the accuracy of the admittance parameters that the user provides, it is important to begin with an accurate representation of the "Device Under Test" (DUT). This modeling process will be demonstrated for a simple circuit board (PDNDUT) that implements a Pi network composed of a single input capacitor, a series resistor with parasitic inductance, and a single output capacitor.

## Capacitor Models

The increasing availability of RF-aware circuit-analysis tools has motivated many component manufacturers to provide S-parameter model libraries for common components such as capacitors. With short-duration development schedules, there is frequently a strong motivation to acquire and use vendor-supplied models with little or no model validation. Let the buyer beware! In recent years there has been growing awareness that the effective capacitance of Multi-Layer Ceramic Capacitors (MLCC) exhibits a strong dependence on operating temperature, DC bias-voltage, and AC bias-voltage [8]. In response, some vendors offer models for specific temperatures and DC-bias conditions, or they allow the generation of models customized to specific temperature and DC-bias conditions. So far, few of the vendors account for the magnitude of the AC-bias (ripple voltage) applied to the part. Consider that MLCC components are commonly tested at 0 V DC-bias with an AC test voltage (bias) of 1.0 Vrms (1.414 Vpeak or 2.828

Vp-p). In a typical decoupling application, capacitors operate with 0.8 V to 1.2 V DC-bias and an AC ripple of 10 mVp-p or less.

To understand the consequences of these operating conditions, consider a 100 uF 6.3V X5R ceramic capacitor used on the test board shown in **Figure 9**. The model for this part was specified at 20 °C with 1.2 V DC-bias. Yet this model did not correlate well with measured results. While the model predicted the device capacitance to be 91 uF (at 10 kHz) with 1.2 VDC-bias, the measured average for 10 capacitors was only 53.1 uF for the same DC-bias using a 4294A Precision Impedance Analyzer. The difference is that the model apparently predicted the capacitance at 1 Vrms AC-bias while the measurements were made at 10 mVrms AC-bias.

To corroborate these findings, subsequent measurements were made using an E5061B VNA with the capacitor mounted on a simple test board, as depicted in **Figure 4**. The accuracy of these measurements can be slightly improved by first measuring a reference THRU path with no components attached and then measuring the component mounted on an identical THRU substrate. The effects of substrate parasitic capacitance and inductance are then mathematically removed from the component measurement data using a process referred to as "de-embedding" [9]. Instrument and software companies also offer commercial products to facilitate de-embedding [10, 11, 12].

A comparison of the vendor-supplied capacitor model along with raw and de-embedded VNA measurements is presented in **Figure 5**. Again it may be observed that the capacitor model predicts a capacitance of 91 uF at 10 kHz, while the raw and de-embedded measurements indicate a capacitance of only 57.1 uF. The salient difference is that the model apparently assumed a typical 1 Vrms AC stimulus while the VNA measurements were made with a source power setting of -16 dBm (equivalent to 35.5 mVrms in a 50 ohm environment).

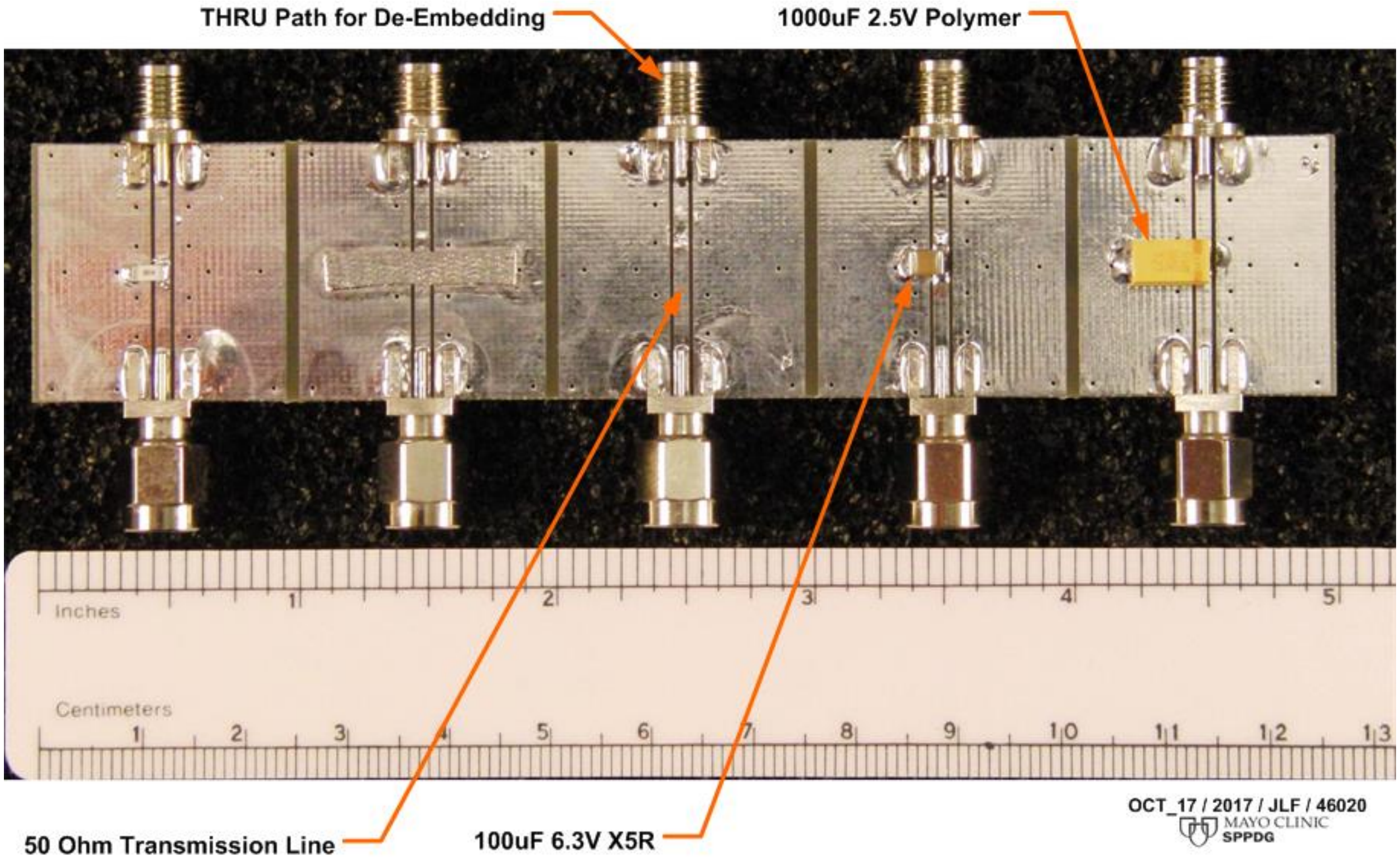

**Figure 4 - Measurement fixture for SMT components**

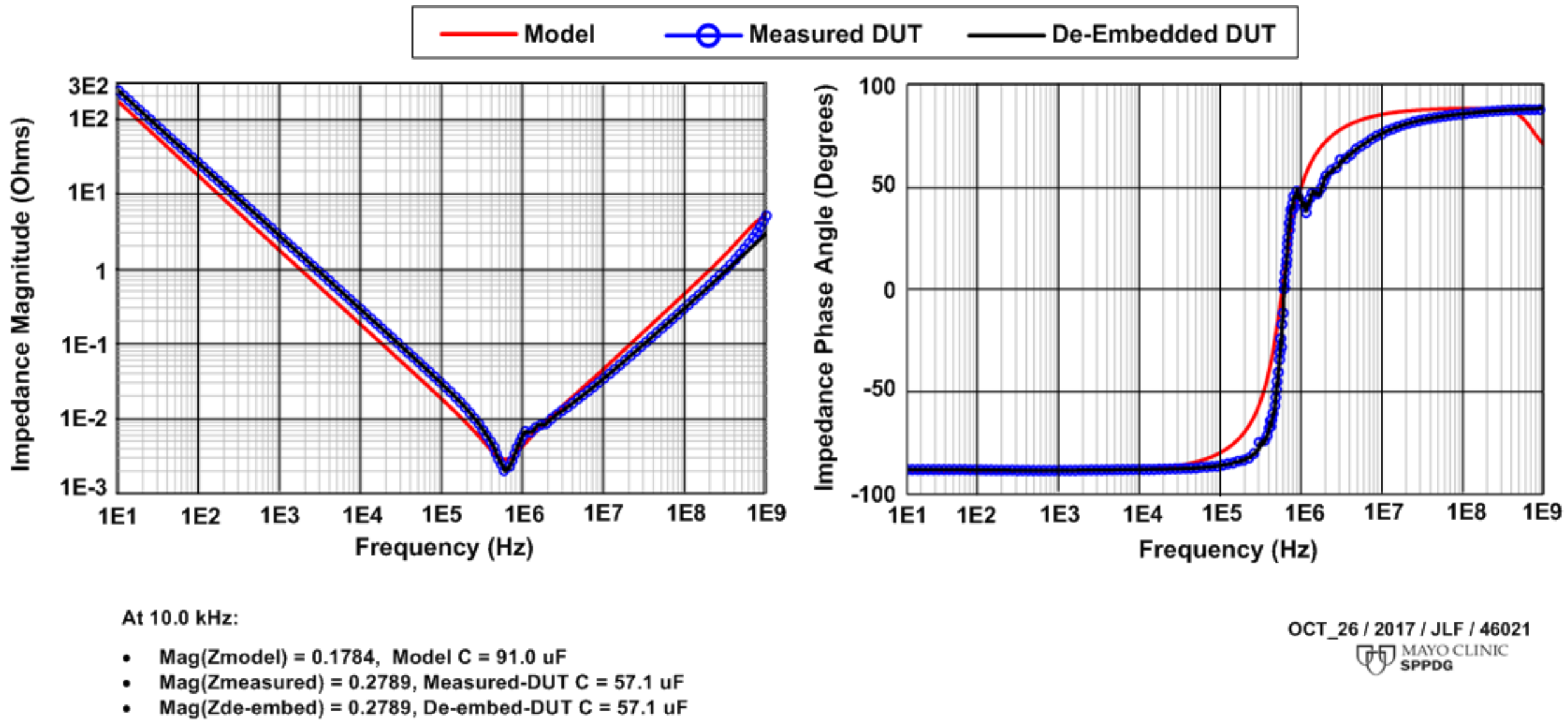

**Figure 5 - Comparison of vendor model to VNA measurements**

The ramifications of this discovery are quite unsettling. It appears that existing vendor models for ceramic capacitors did not accurately predict component behavior for parts in low-voltage DC-supply decoupling applications. The lesson from this experience is quite clear: all component models must be validated by measurements under the same operating conditions (temperature, DC-bias, and AC-bias) that these components are likely to experience in the circuit.

Once the component measurements are complete they should be examined for any anomalies. Measurement fixtures and/or network analyzers may introduce slight errors that can complicate subsequent NICE steps. One way to mitigate those errors is to derive an idealized lumped-element equivalent-circuit that matches the S-parameter response of the original component. This step is easily accomplished using an RF-aware circuit simulator that supports optimization. The process starts with a single series-RLC branch that approximates the low-frequency response of the capacitor of interest. Additional series-RLC branches are then added with progressively smaller inductance and capacitance values to match the response at middle and upper frequencies. An example of such a simulation is illustrated in **Figure 6**, while the results are depicted in **Figure 7**. Experience suggests that a typical MLCC may require up to six branches to achieve a satisfactory match. Additional information regarding model optimization is available online [13, 14, 15].

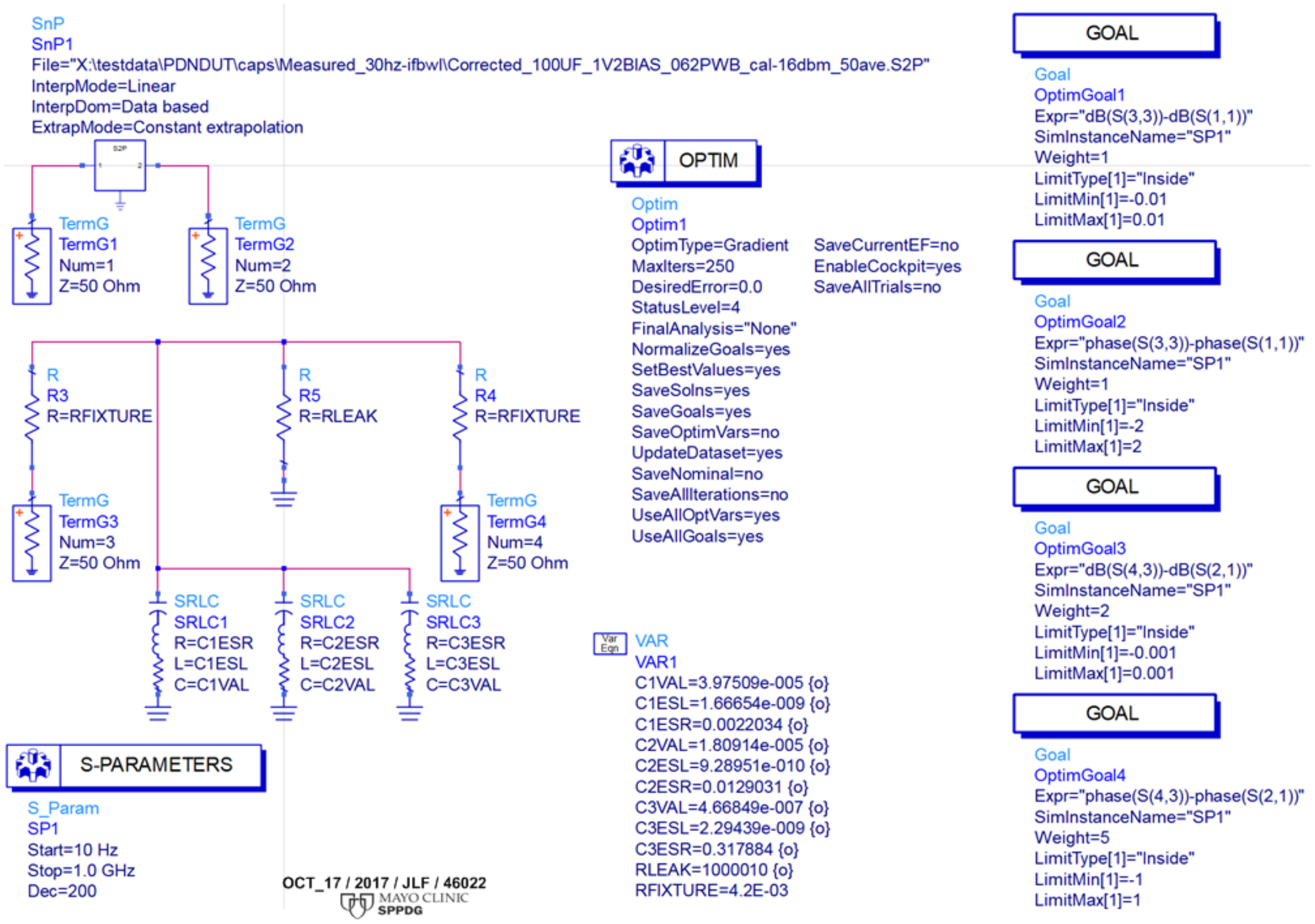

**Figure 6 - Optimization of equivalent circuit for 100uF X5R capacitor**

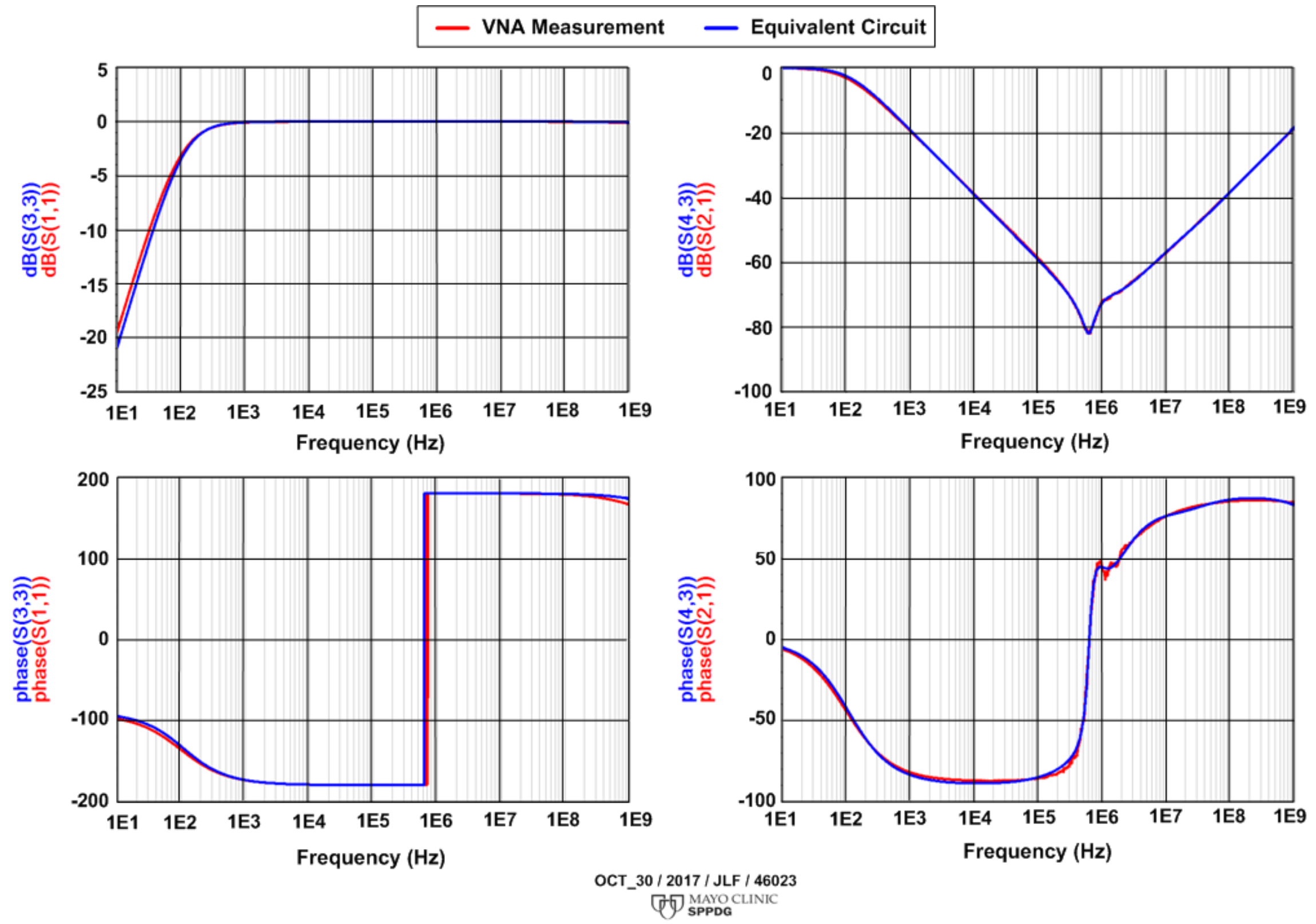

**Figure 7 - Comparison of measured and equivalent-circuit S-parameters for ceramic capacitor model**

## Board Models

Once the component models are complete, it is necessary to create a model for the complete PDN, including the effects due to the printed circuit board itself. This task can be accomplished by the use of a commercial electromagnetic analysis tool. This approach is demonstrated with a simple example: Consider the circuit shown in **Figure 8** from the PDNDUT board.

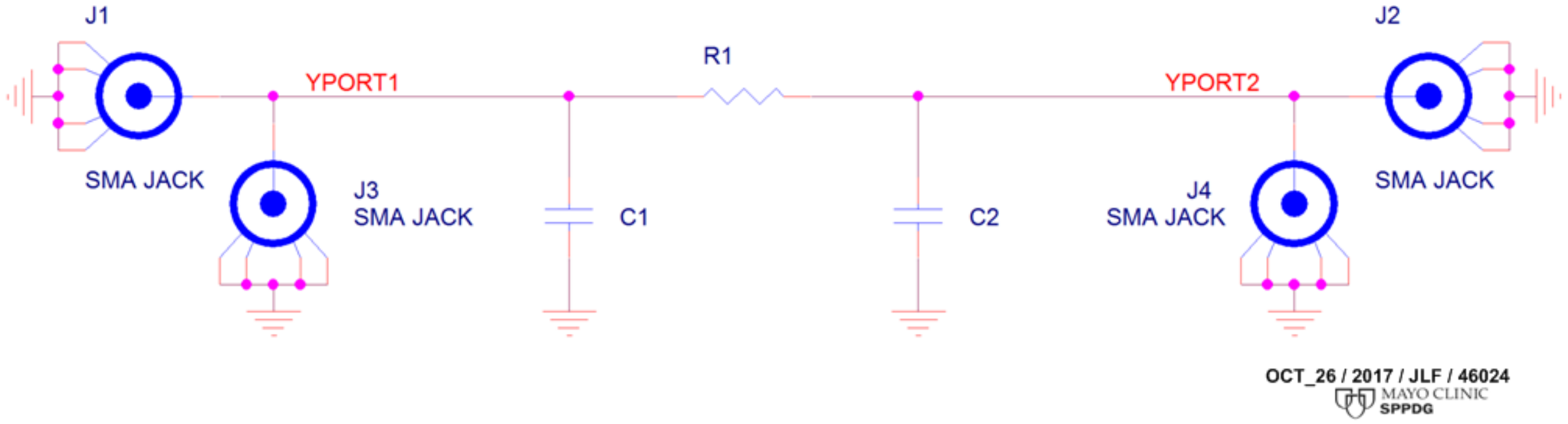

**Figure 8 - Circuit diagram for PDNDUT board**

The circuit consisting of signals YPORT1 and YPORT2 implements the classic Pi network shown in **Figure 2**. Connectors J1 and J2 represent the "thru" path of a PDN while connectors J3 and J4 provide convenient connections to monitor the voltages at J1 and J2, respectively. For this example, the board was assembled as depicted in **Figure 9**.

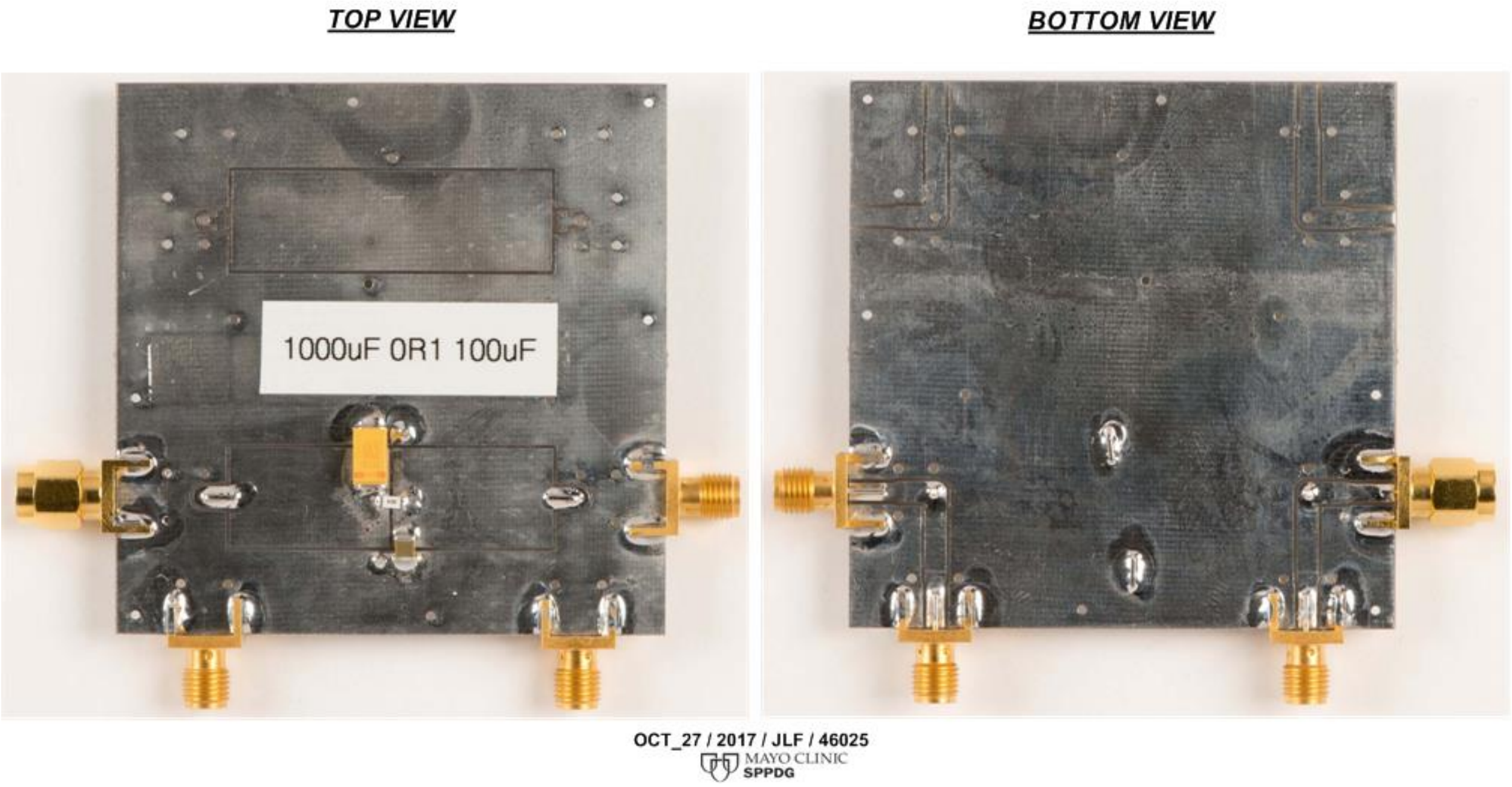

**Figure 9 - PDNDUT board with components**

By careful adjustment of simulation parameters such as dielectric permittivity and trace conductivity, it was possible to achieve correlation of simulated and measured network parameters for the PDNDUT board with only R1 installed, as illustrated in **Figure 10**.

For validation of the complete PDNDUT assembly, a circuit simulation was performed, as shown in **Figure 11**, to facilitate display and comparison of the relevant parameters. Recall that the expression for the network output current I2 was given by equation (2):

$$\boldsymbol{I2 = Y21 * V1 + Y22 * V2}$$

Consequently, final model validation should focus on network transfer admittance Y21 and network output admittance Y22. The corresponding admittance parameters of interest listed in **Table 1** were derived using the simulation configuration shown in **Figure 11**.

| Network Parameter | Case: J3 & J4 Open | | Case: J3 & J4 Terminated | |
|---|---|---|---|---|
| | Measured | Simulated | Measured | Simulated |
| Network transfer admittance | Y(2,1) | Y(4,3) | Y(6,5) | Y(8,7) |
| Network output admittance | Y(2,2) | Y(4,4) | Y(6,6) | Y(8,8) |

**Table 1 - Network Parameter vs. Simulation Quantity**

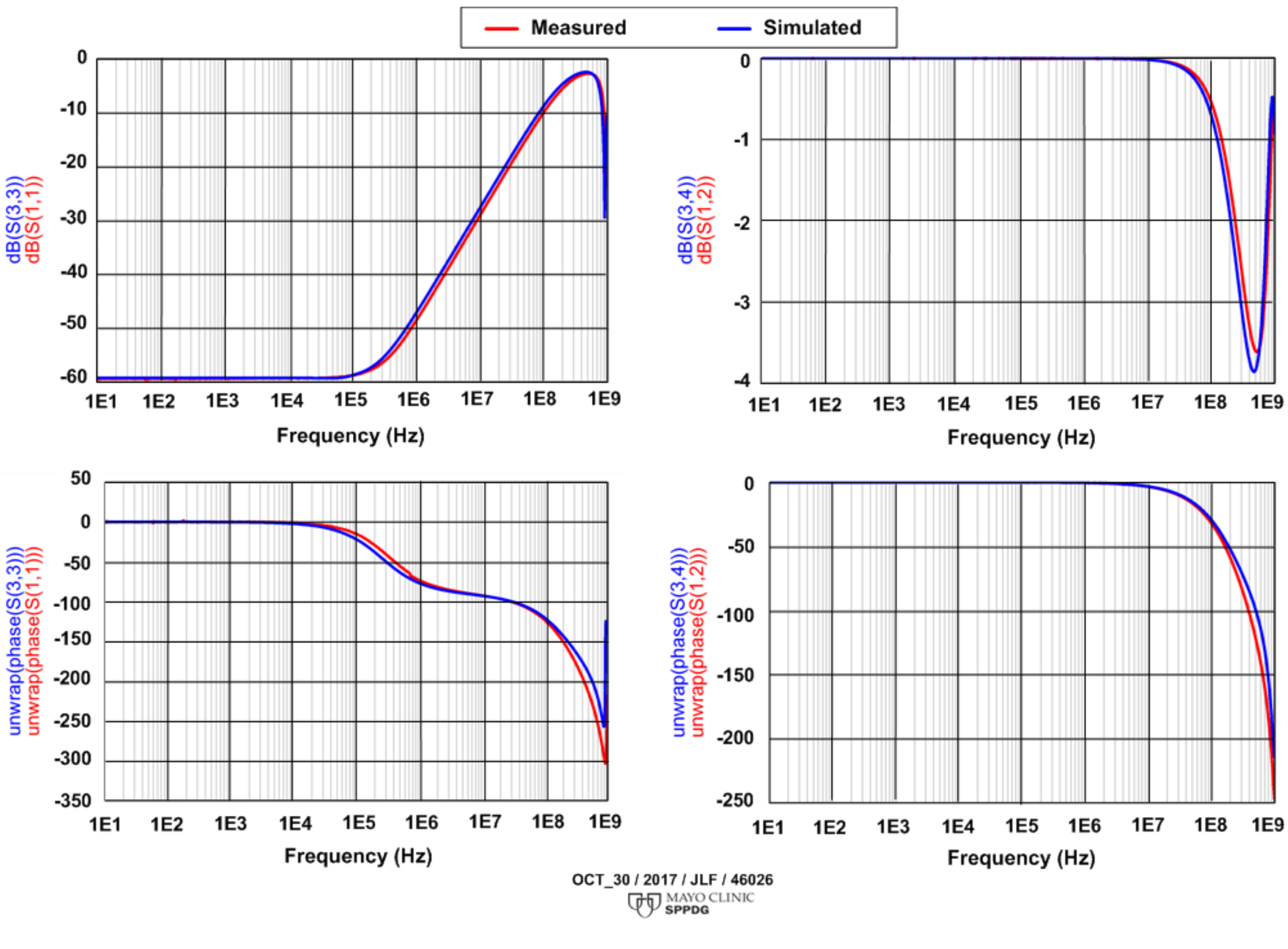

**Figure 10 – S-parameters for PDNDUT J1, J2 with J3, J4, C1, and C2 open**

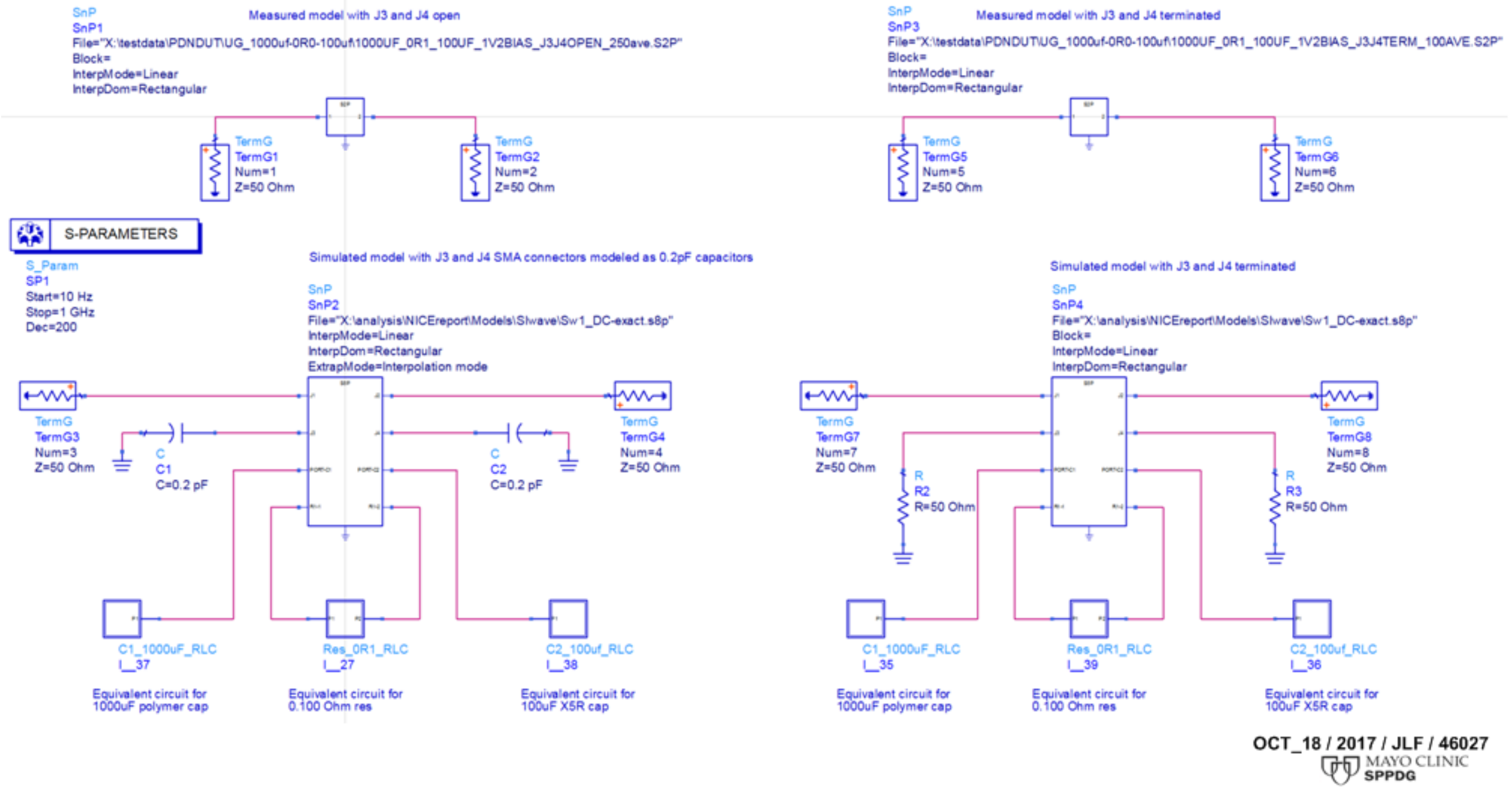

**Figure 11 – Comparison of measured and simulated network parameters for PDNDUT assembly**

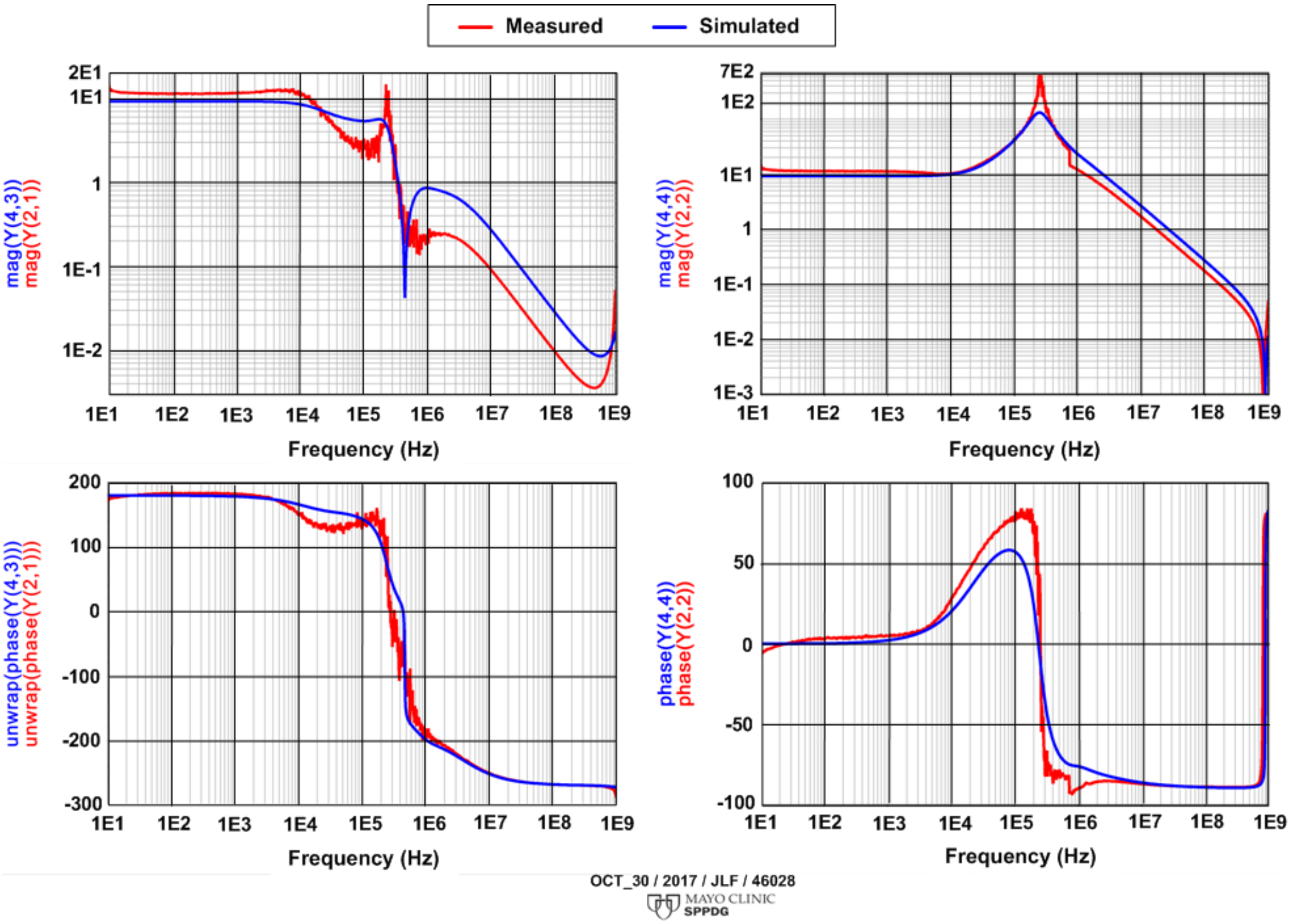

**Figure 12 - Measured and simulated transfer and output admittances for PDNDUT with J3, J4 open**

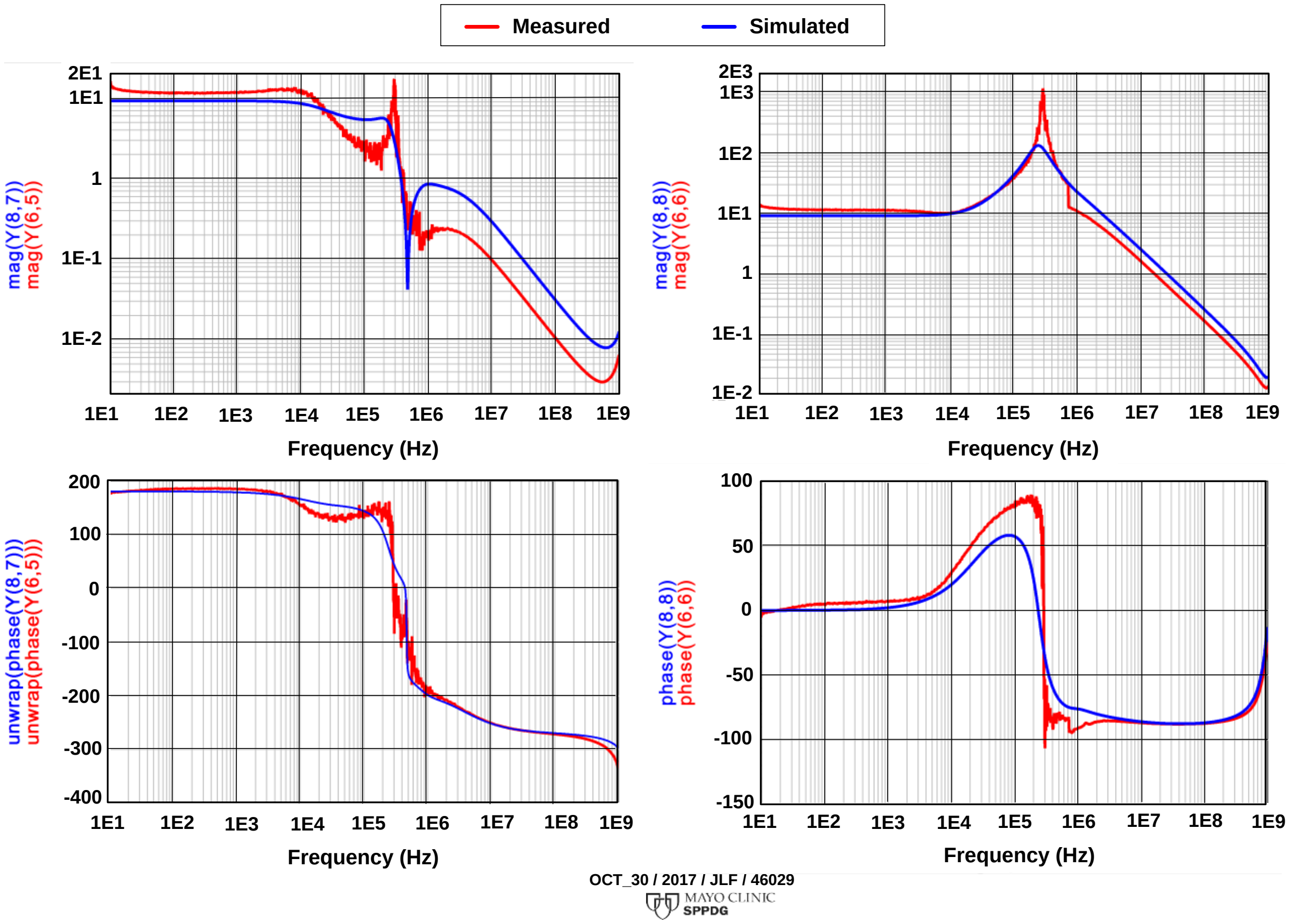

**Figure 13 - Measured and simulated transfer and output admittances for PDNDUT with J3, J4 terminated**

The results detailed in **Figure 12** and **Figure 13** show reasonable correlation between simulated and measured admittance parameters except for two conditions:

1. Near the circuit-resonance (250 kHz), the simulated admittance magnitudes and phases varied from the measured values in all cases. Fortunately, the deviation of the output admittance is restricted to a fairly narrow frequency range, so the effect on the convolution of the voltage waveforms is presumed to be small. No definitive cause for the transfer admittance deviation below the resonance has been found, though there is speculation that this may be a consequence of the tin-lead Hot-Air Solder Leveling (HASL) finish applied to the PDNDUT boards. The planar EM solvers are incapable of solving for fields within the metal conductors and consequently, these tools are unable to model copper with a thin layer of tin-lead applied to the surface. Since the conductivity of copper is roughly eight times greater than tin-lead, it is not unreasonable to expect that HASL boards would exhibit skin-effect consequences that would not be included in an EM-extraction for traces modeled as copper-only. We plan to order any future test boards with “bare copper” surface finish to mitigate this concern.

2. At 700 kHz, a significant step reduction is observed in the measured magnitudes, with a fixed offset of nearly a half an order-of-magnitude for the remainder of the sweep. We believe that this step is an artifact of the E5061B VNA used for this measurement. The E5061B uses multiple receivers to span the frequency range from 10 Hz to 1 GHz, and it switches from one receiver to another at 700 kHz. While this switch operation does not upset the S-parameter measurements substantially, the effect on the Y-parameters is large. Unfortunately, until a remedy can be found, this apparent error in the measured admittance parameters limits the utility of measured models for use in NICE measurements and emphasizes the need to construct accurate EM-simulation and component models.

So far, admittance parameters have been examined for the case where J3 and J4 are open, as well as the case where J3 and J4 are terminated. For the NICE voltage-measurement setup on PDNDUT, both J3 and J4 were connected to Keysight N7020A "Power Rail" probes. Figure 6 of the N7020A User's Guide [16] shows that the input impedance of the probe varies from approximately 50 kohms at low frequencies to 50 ohms at high frequencies. To maximize the accuracy of the PDNDUT network parameters, effects from probe-impedance variation were considered.

Figure 18 of the N7020A User's Guide [16] presents an equivalent circuit for the probe composed predominately from transmission-line segments. While this model is useful for simulations, the author's experience using this model was that the simulations ran slowly and the results tended to show the same kind of low-magnitude variation above 100 MHz that was observed in the input impedance (Figure 6 of [16] discussed above).

To reduce simulation time and mitigate the variations in the high-frequency response, the E5061B was used to make a one-port measurement of the probe input return-loss S(1,1) and the ADS optimizer was used to find component values for a lumped-element model that matched the measured S(1,1). The resulting lumped-element equivalent-circuit is illustrated in **Figure 14**, and the resulting input impedance is displayed in **Figure 15**. The impedance was observed to be free from variations at high frequencies.

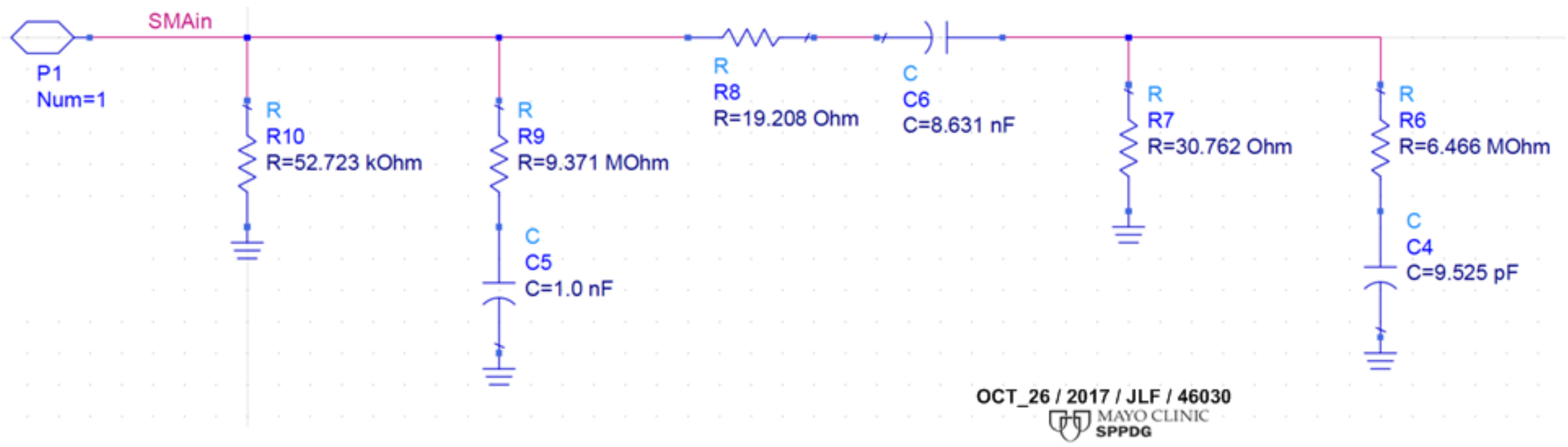

**Figure 14 – Simplified equivalent circuit for N7020A probe input**

The simplified probe model allowed the creation of an overall model for the PDNDUT board assembly and N7020A probes, as shown in **Figure 16**. A simulation of this model produced the admittance parameters shown in **Figure 17**. These Y-parameters were then exported in Touchstone® format for use during the NICE measurement.

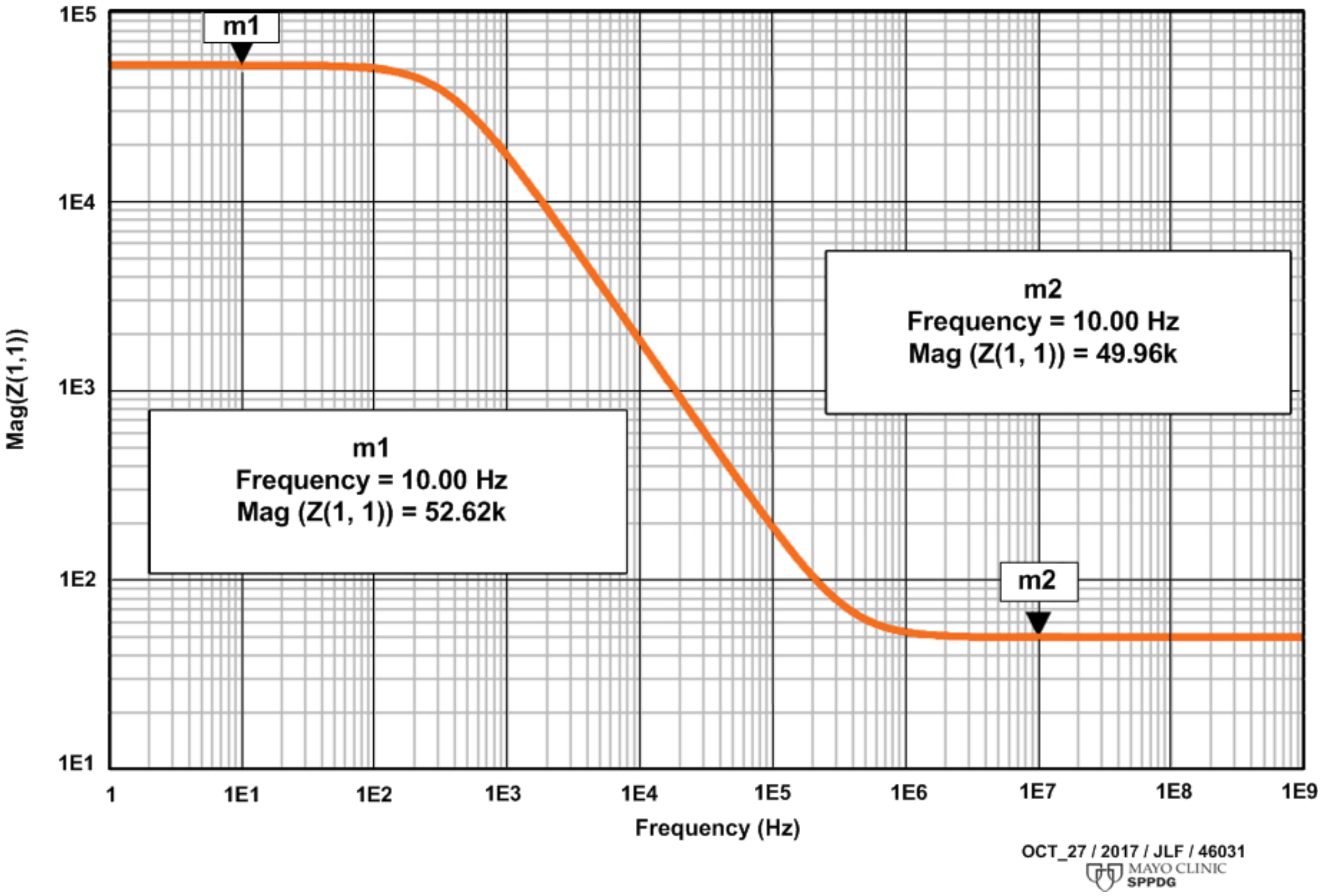

**Figure 15 - Simulated input impedance of simplified N7020 probe model**

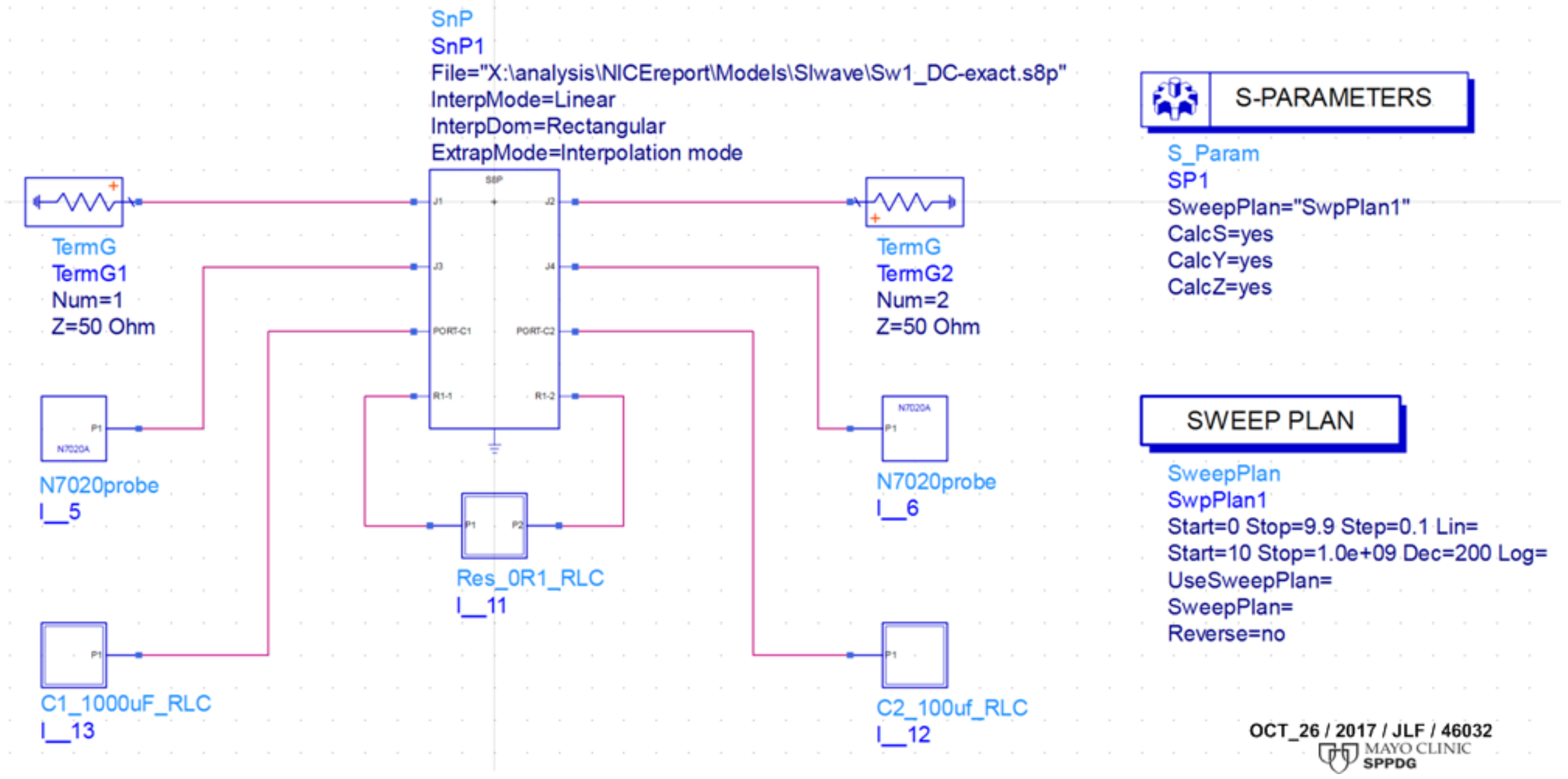

**Figure 16 - Overall model for PDNDUT assembly with N7020A probes**

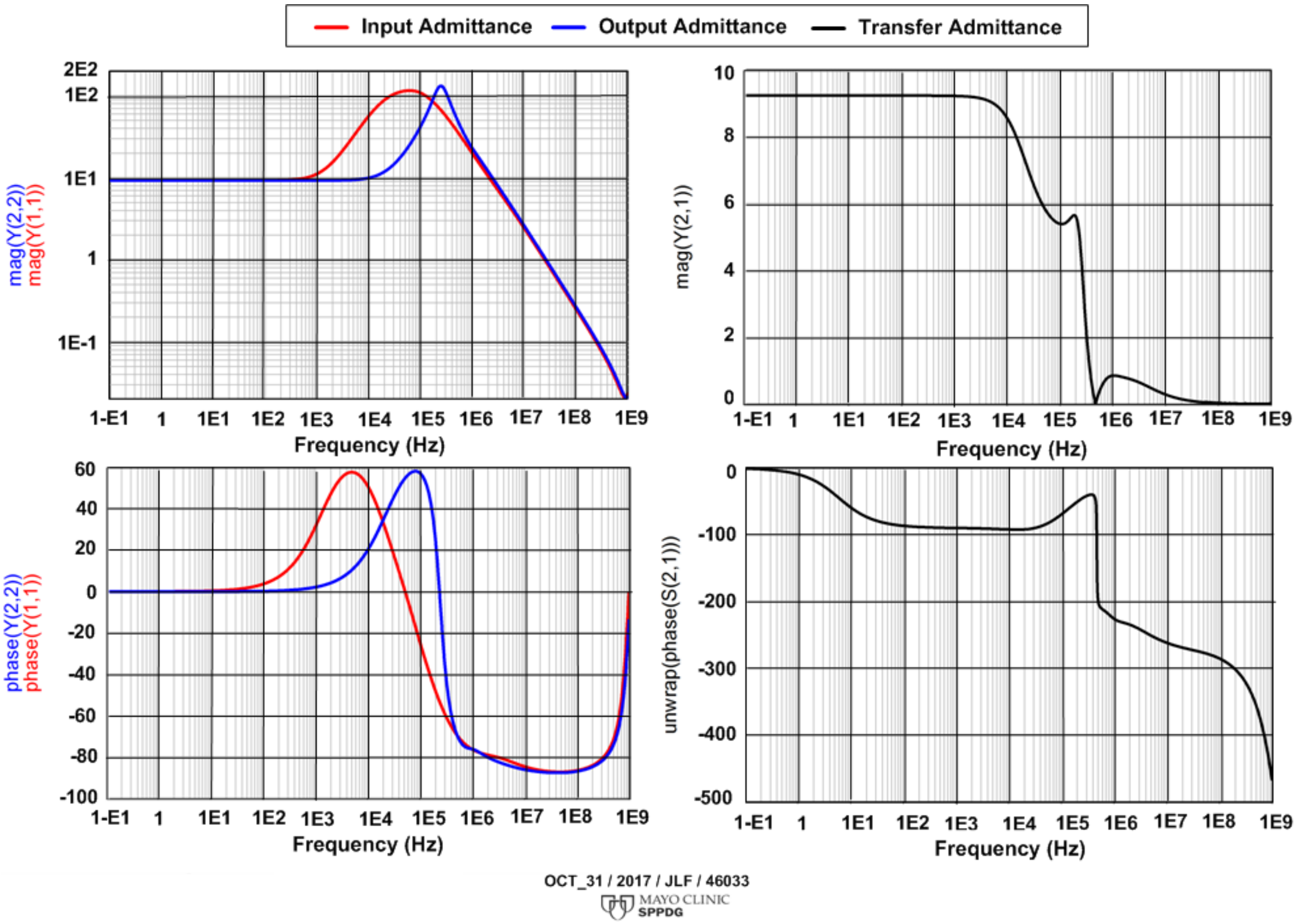

**Figure 17 - Simulated Y-parameters for PDNDUT with N7020A probes**

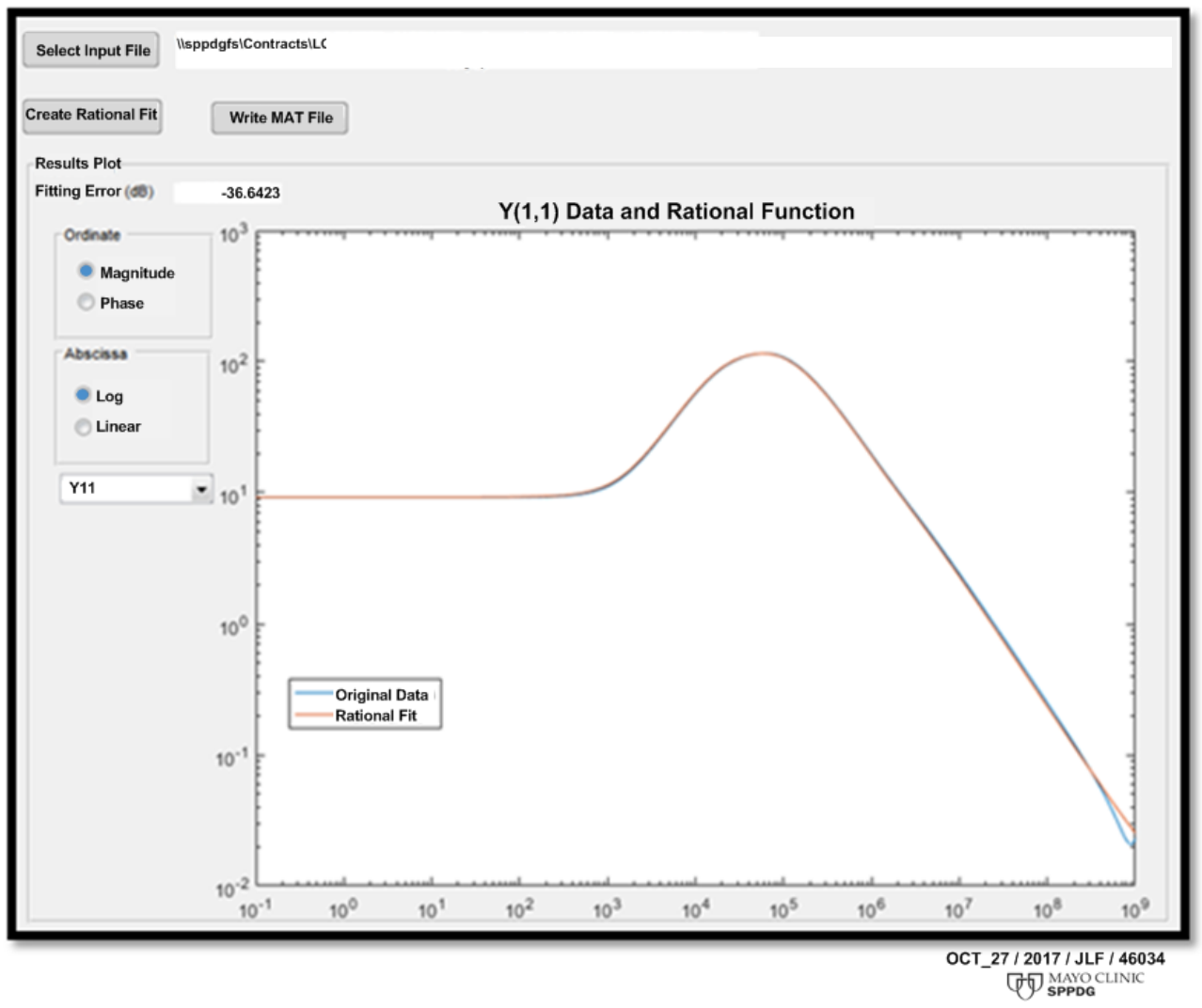

**Figure 18 - GUI for generating the rational-fit file**

The last preparatory step was the creation of a “rational-fit file” whereby the tabulated network parameters are adapted to a pole-residue format before use in the NICE-waveform processing. This step accelerates calculation of the current waveforms once the voltage waveforms have been captured.  To simplify this process, a custom Matlab® GUI was written.  The graphical output from the rational-fit generation appears in **Figure 18**. For this example, it is observed that the rational-fit provides excellent correlation to the original admittance parameters up to approximately 300 MHz. Since a typical PDN is deliberately designed to suppress high-frequency ripple, it was logical to assume that such minor discrepancies at high frequencies between the measured Y-parameters and the rational-fit are negligible. The rational-fit file was then saved on the oscilloscope for subsequent use by the waveform processing routine.

# NICE Measurements

As illustrated in **Figure 19**, the general case of NICE measurement requires only two connections to the device or system under test:

1. A connection to the power supply net of interest (and ground plane) along the current path near the Voltage Regulator Module (VRM).
2. A connection to the same power supply net (and ground) along the current path near the load. Preferably this connection is made using voltage and ground (VDD & VSS) sense signals originating from inside the ASIC. These signals are typically provided by ASIC designers to provide accurate feedback to the voltage regulator. If such sense signals are not available then the supply and ground nets on the board near the ASIC would be probed.

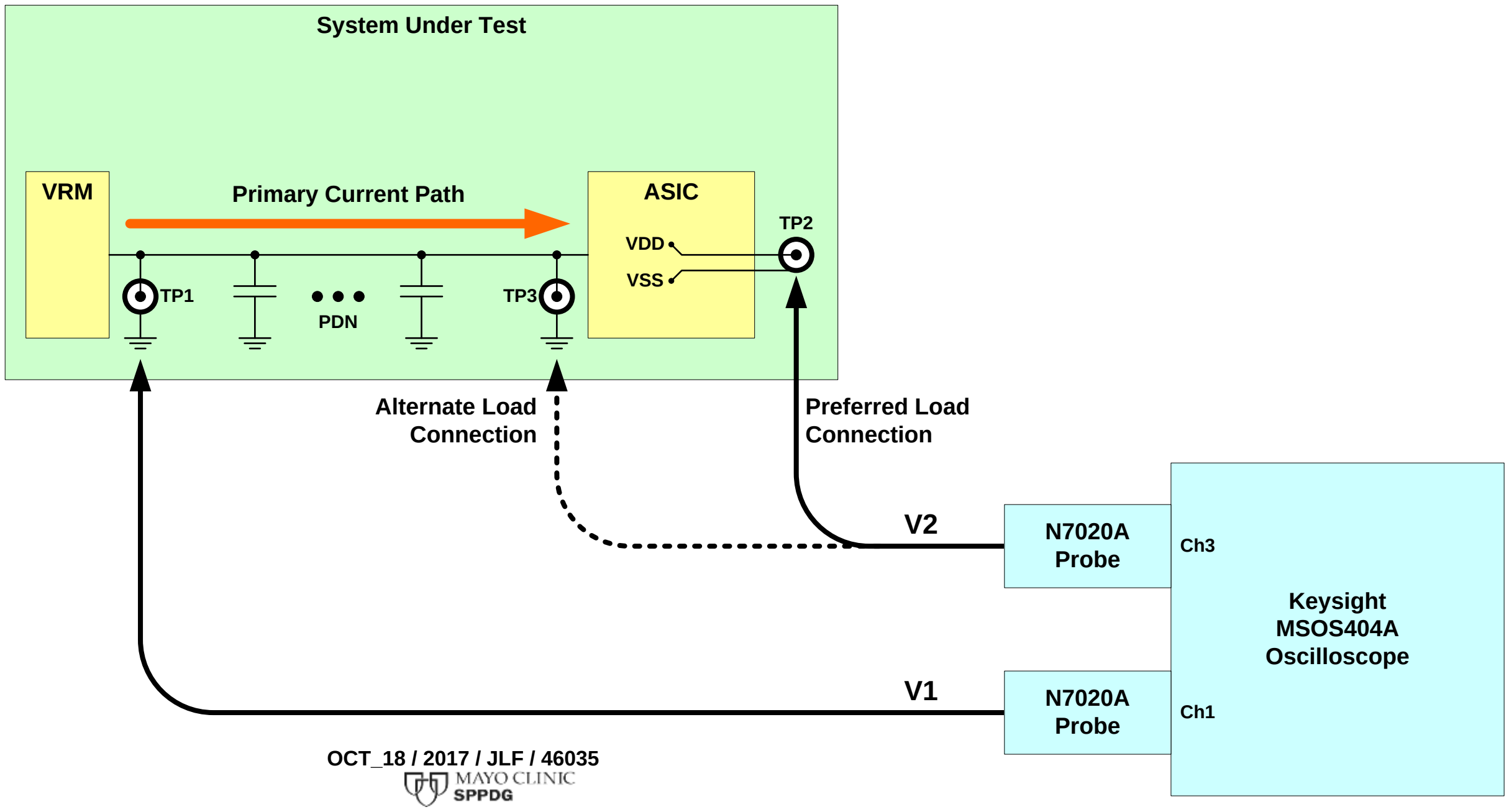

**Figure 19- Required connections for making a NICE measurement**

To demonstrate a NICE measurement using the PDNDUT card, additional external equipment was necessary to perform the functions provided by the voltage regulator (1.2V supply) and the load (a Picotest J2112A controlled current-sink driven by an arbitrary-waveform generator), as depicted in **Figure 20**. The Picotest J2112A also includes a convenient “Monitor” port providing access to a voltage proportional to the J2112A output current, which is useful for validating the NICE current.

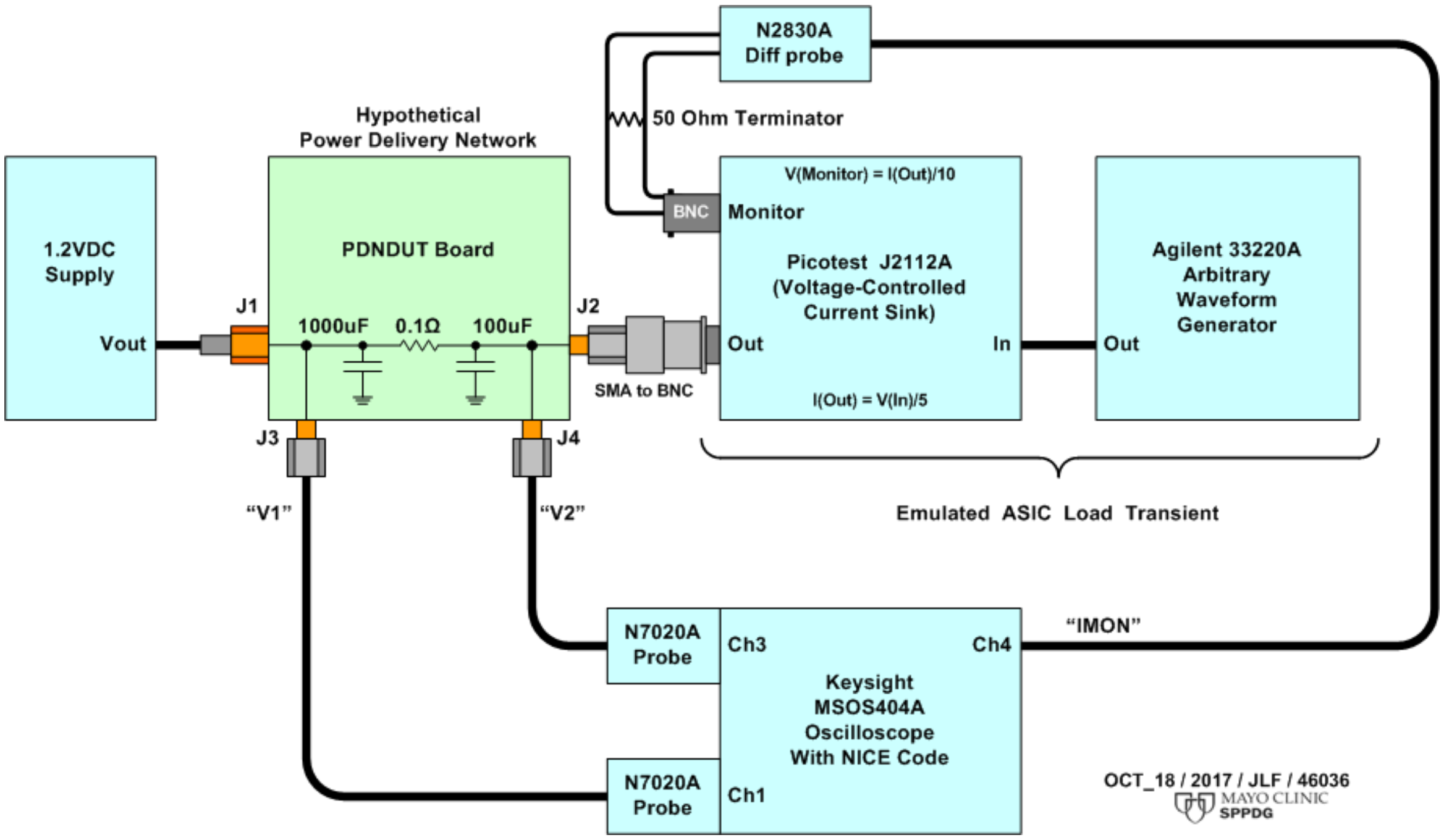

**Figure 20 - Equipment setup for NICE validation**

In this example, the oscilloscope was configured to trigger on the falling edge of V2 (the supply voltage at the load). Once waveforms were captured for V1 and V2, a trace was added to display the calculated NICE current using the add-trace function, as illustrated in **Figure 21**. After a short delay for processing, the NICE current waveform was displayed on the oscilloscope as presented in **Figure 22**.

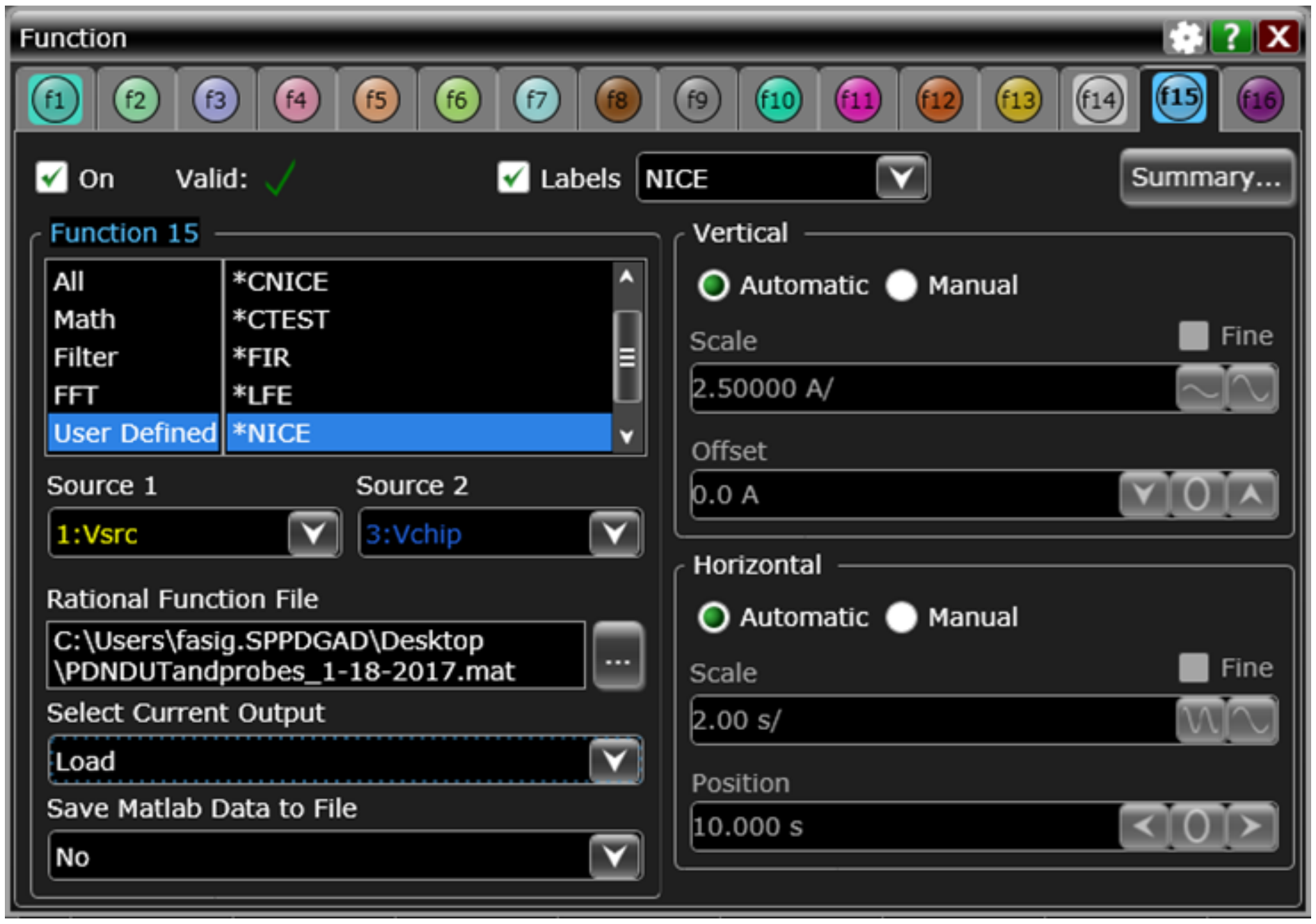

**Figure 21 – NICE function settings**

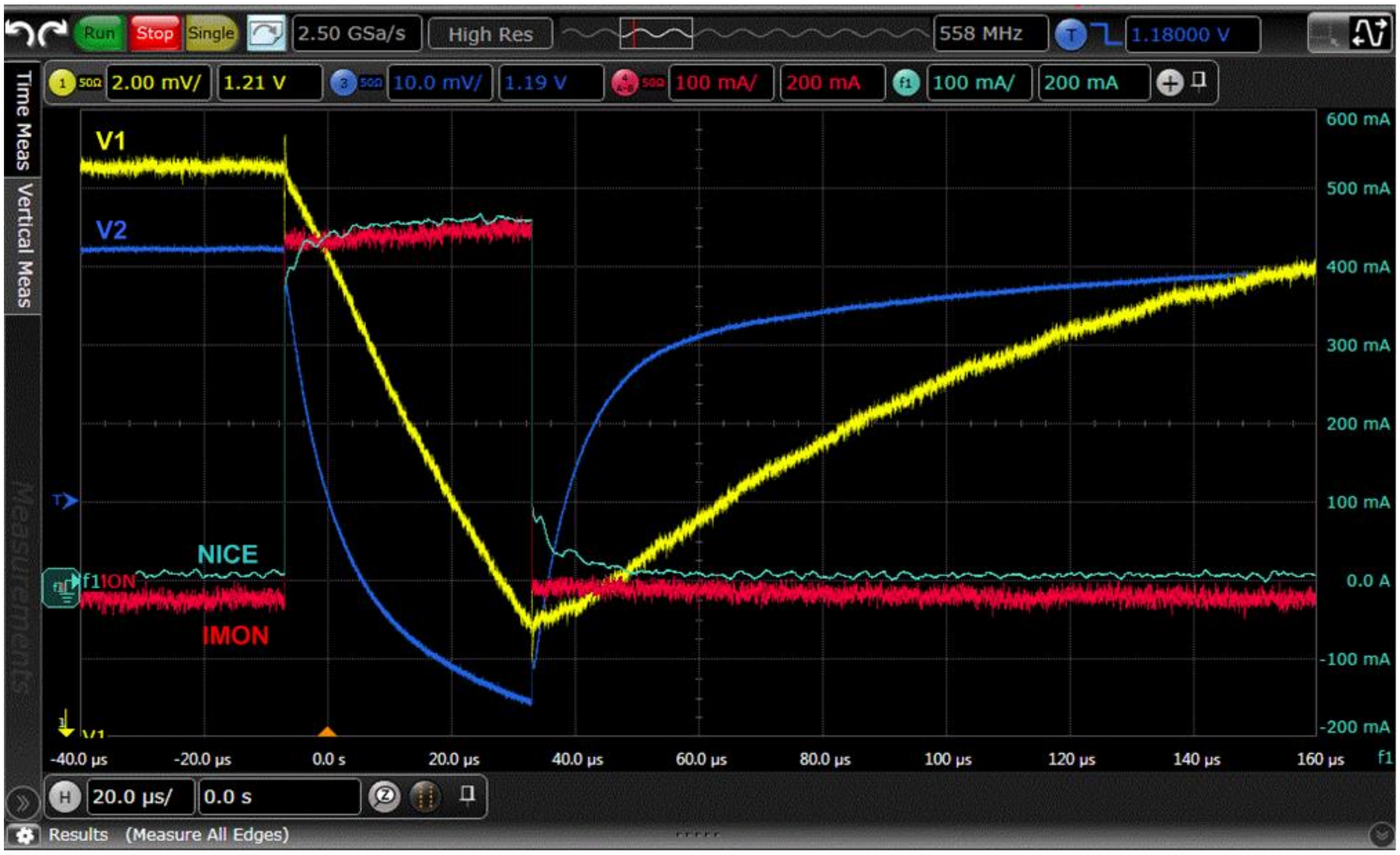

OCT_18 / 2017 / JLF / 46038
MAYO CLINIC SPPDG

**Figure 22 - Oscilloscope display of PDNDUT Monitor (IMON) and NICE currents**

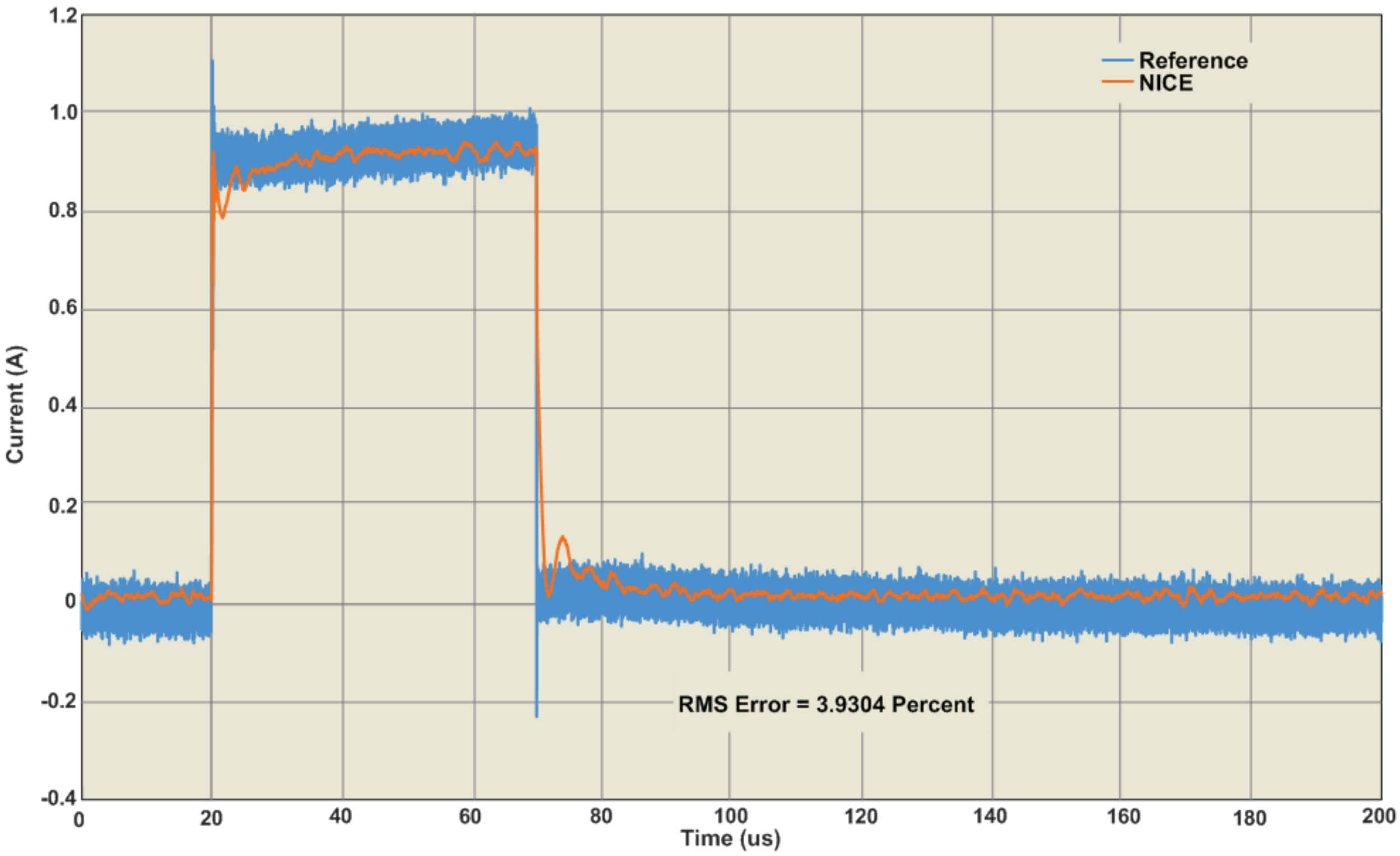

OCT_18 / 2017 / JLF / 46039
MAYO CLINIC SPPDG

**Figure 23 - Comparison of reference and NICE currents for pulse-load with 10 ns rise time**

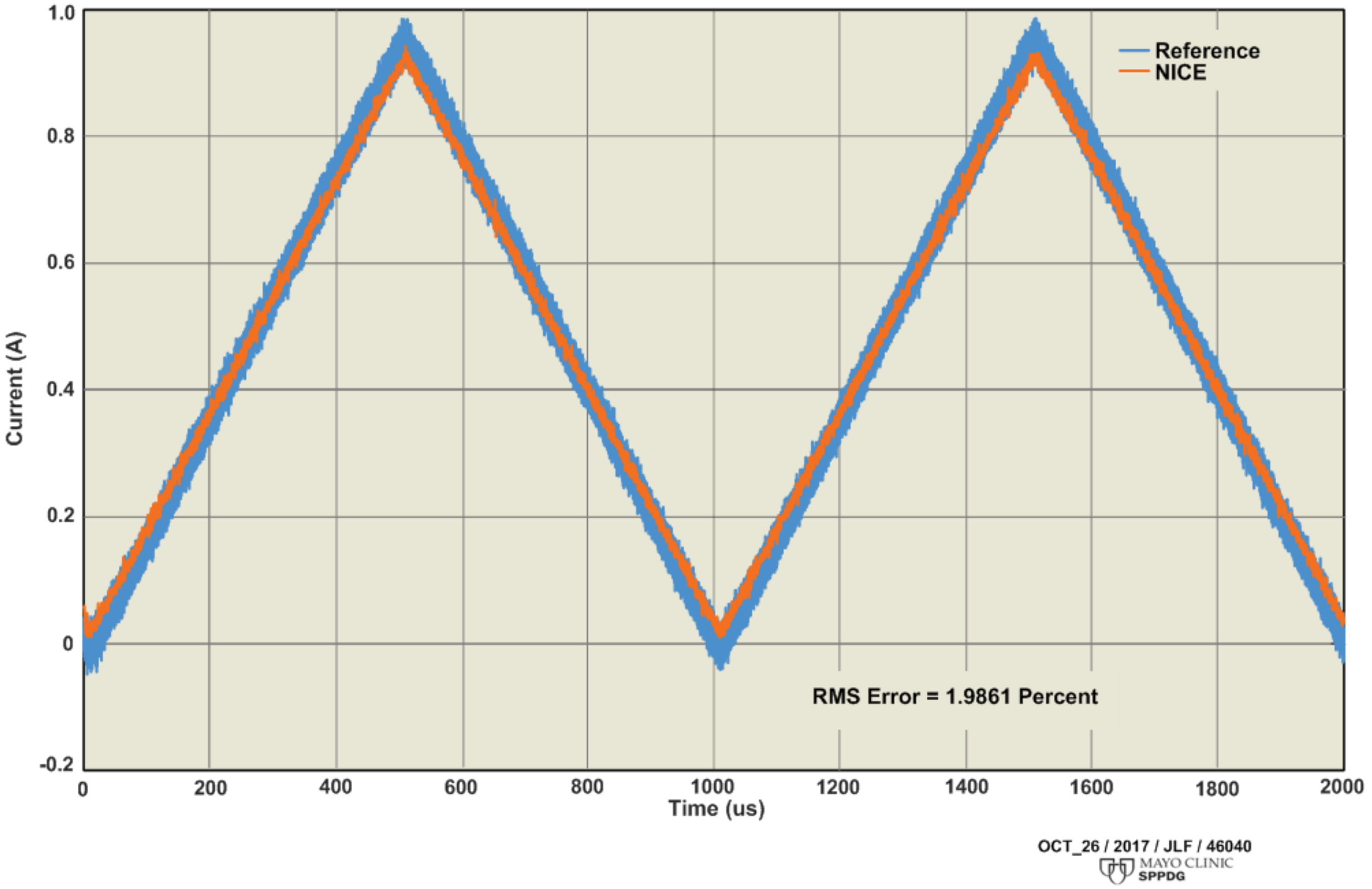

**Figure 24 - Comparison of reference and NICE currents for a linear ramp load**

**Figure 23** and **Figure 24** illustrate point-by-point comparisons of the NICE waveform and reference (IMON) waveform. The RMS error in the NICE current is less than four percent. Similar analysis indicates that comparable errors result for sine, sawtooth, and exponential load-current profiles.

## NICE Algorithm

The deployment of NICE was done directly on the oscilloscope utilizing the User Defined Function (UDF) interface. The algorithm to produce a NICE output from input measurements can be divided into three key phases.

The first phase of the algorithm is the collection of modeled or measured admittance parameters in the frequency domain. However, the goal is to generate time domain current waveforms from time domain voltage inputs. This step can be conducted in several different ways. One solution is to convert the time domain inputs to the frequency domain for processing with the DUT data, and then convert the outputs to the time domain, as illustrated in **Figure 3**. Another solution is to convert the frequency domain DUT data to the time domain and conduct the operations in the time domain. For this example, the latter was chosen.

In the second phase of the algorithm, the frequency domain representation of the DUT is converted to a time domain representation. The frequency domain to time domain conversion is accomplished using vector fitting [18]. In this process, the frequency domain data is approximated with a pole-residue function. This approach has a number of benefits. First, it guarantees that the time domain solution is causal.  Second, it is possible to test and correct network passivity [19].  Finally, it is possible to convert the solution into a state space representation [20].

In the final phase of this algorithm, the time domain representation of the DUT is used to produce time domain results.  The time domain work is accomplished using state space methods.  The two voltages of interest are measured using the approaches previously discussed.  The waveforms are then passed as the inputs to a state space model that has been constructed with the DUT time domain representation.  The output of the state space model is the currents at both ports of the DUT.  This approach is much faster than the equivalent frequency-time-frequency transform processes, and it is possible to process the waveforms in near real-time.

## NICE With Multi-Pin Devices

There is no simple means to directly measure the total instantaneous supply current entering a multi-pin component like an FPGA, which is why the NICE method was developed.  Consequently, there is no direct measurement that can be compared to the NICE waveform to demonstrate NICE accuracy and validity.

Recall from **Figure 1** that two-port network theory allows us to calculate the current entering both ports of the network, and the NICE algorithm actually calculates both I1 and I2 waveforms.  Consider the circuit shown in **Figure 25**.  The Intel/Altera A10DK board has a large multi-pin FPGA that we can configure for various load-current profiles, and it has a PDN which we can model and for which we can compute the relevant admittance parameters.  This particular board also has a series-connected sense resistor between the regulator and the decoupling network.  Therefore Ireg can be computed from the ratio of Vsens/Rsens, and this measured Ireg can be compared to the NICE I1.  From Kirchhoff's Current Law the argument can be made that correlation of measured I1 and NICE I1 suggests that the NICE I2 is also a reasonable estimate.  Therefore this approach provides a validity theorem in place of a conclusive validity proof.

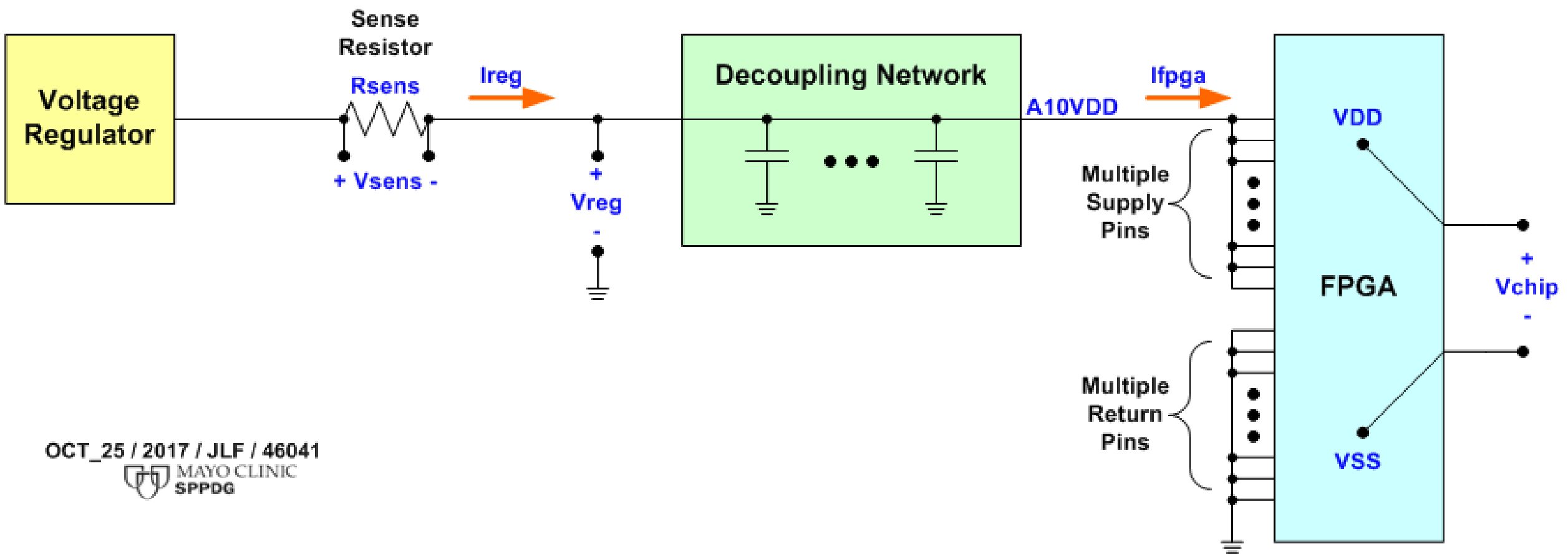

**Figure 25 – Simplified core-logic PDN on the Intel/Altera A10DK board**

The results from this approach are depicted in **Figure 26**.  Approximately 9.56 milliseconds after the FPGA configuration is complete (CONFIG_DONE goes true) the supply voltage at the device (Vchip) begins to fall due to increasing FPGA current.  The regulator responds by raising its output (Vreg) to compensate for I*R voltage drops in the supply conductors.  Initial current demands are supplied by the decoupling network and are replenished by increased regulator current (Ireg).

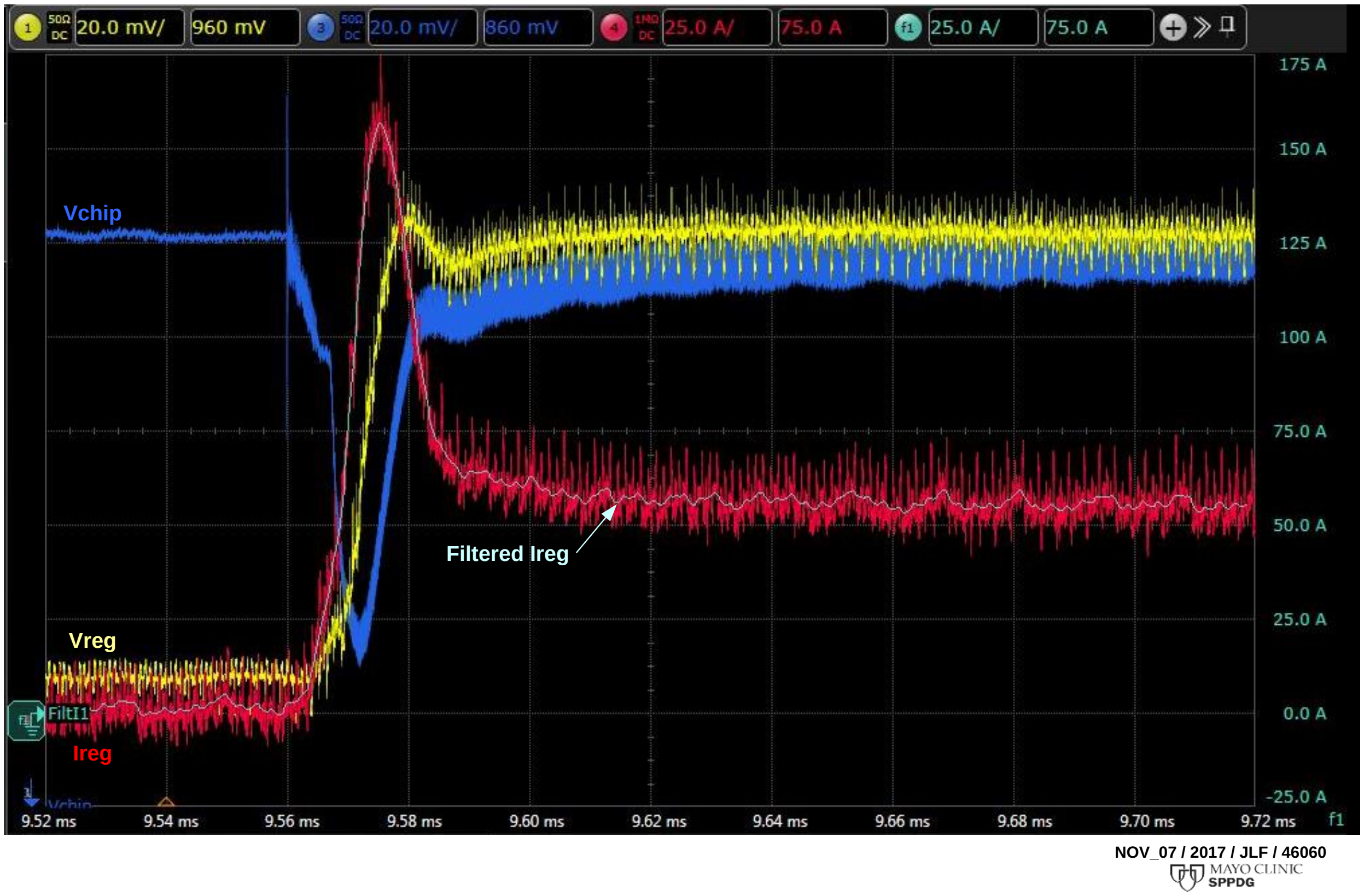

**Figure 26 - Measured Input Voltage, Output Voltage, and Regulator Current on A10DK board**

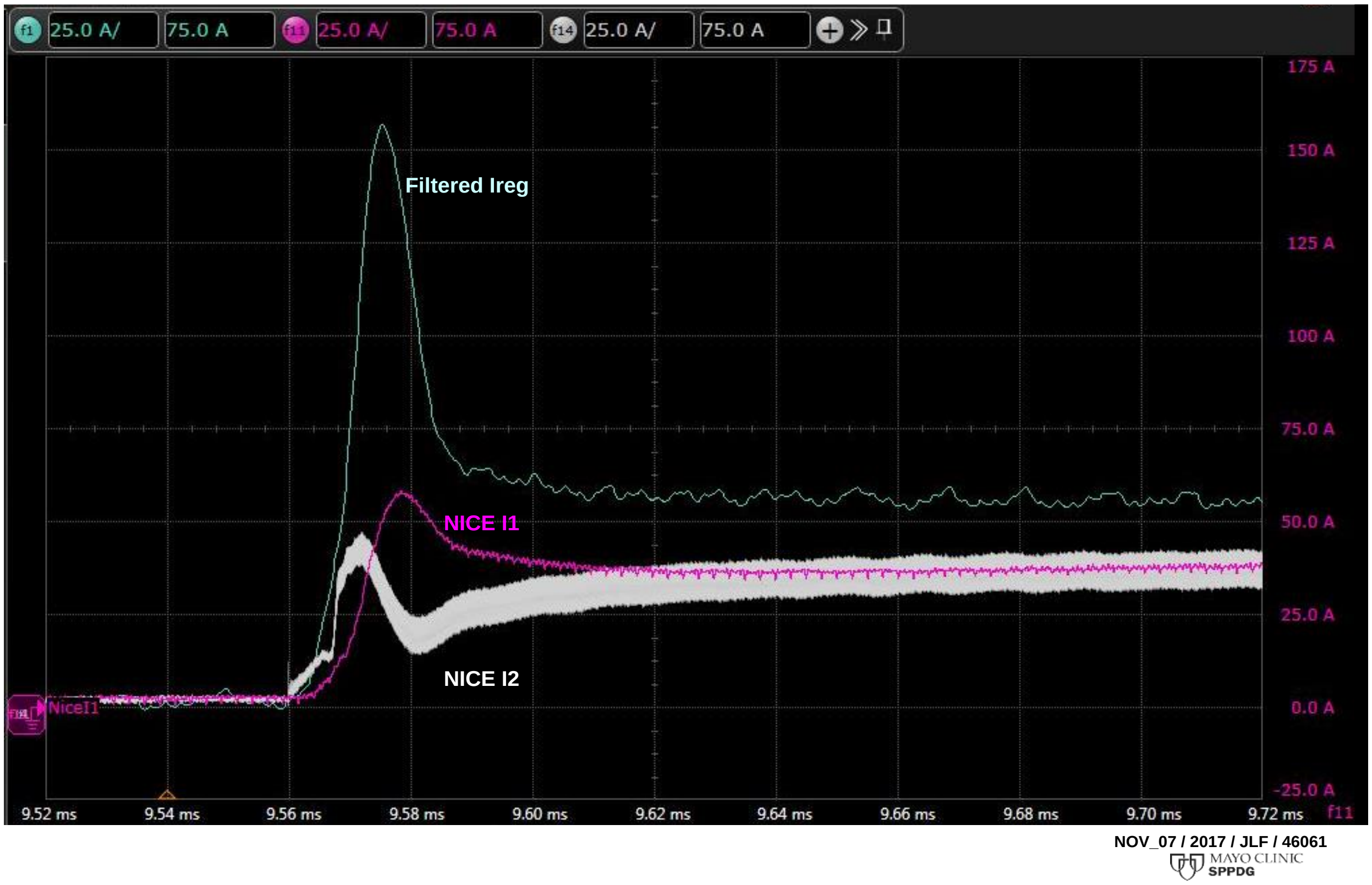

**Figure 27 - Regulator Current and NICE Estimate for PDN Input Current**

Subsequent filtering of Ireg results in the plot of regulator current (Filtered Ireg) displayed in **Figure 27** along with the NICE estimate for PDN input current (NICE I1) and FPGA current (NICE I2). At first glance, the correlation between Filtered Ireg and NICE I1 is not striking. We believe this disparity may be related to the package model for the FPGA. Clearly, resolution of this issue will require additional study and that effort is ongoing.

Despite this disparity, useful observations can be made from the current estimates presented in **Figure 28**. Note that the regulator current (NICE I1) and FPGA load current (NICE I2) begin at the same level, consistent with the FPGA in a low-power, un-configured (idle) state, and the two currents converge to the same steady-state value as expected once the transient recovery is complete and the decoupling network has been replenished. Also note that the increasing FPGA load current mirrors the decreasing supply voltage at the load (Vchip) due to I*R losses in the board and package conductors.

This insight into the FPGA load-current waveform enables speculation regarding the process of FPGA configuration and startup. The interval from 9.560 ms to 9.566 ms likely corresponds to the on-chip clock generators stabilizing at the respective programmed frequencies. Once the clock generators have stabilized, the clock output-buffers are enabled and the logic nodes begin switching resulting in the sharp load-current increase between 9.566 ms and 9.570 ms. This load-current rise and the resulting supply-voltage drop are the motivation for careful and thorough decoupling-network design and validation.

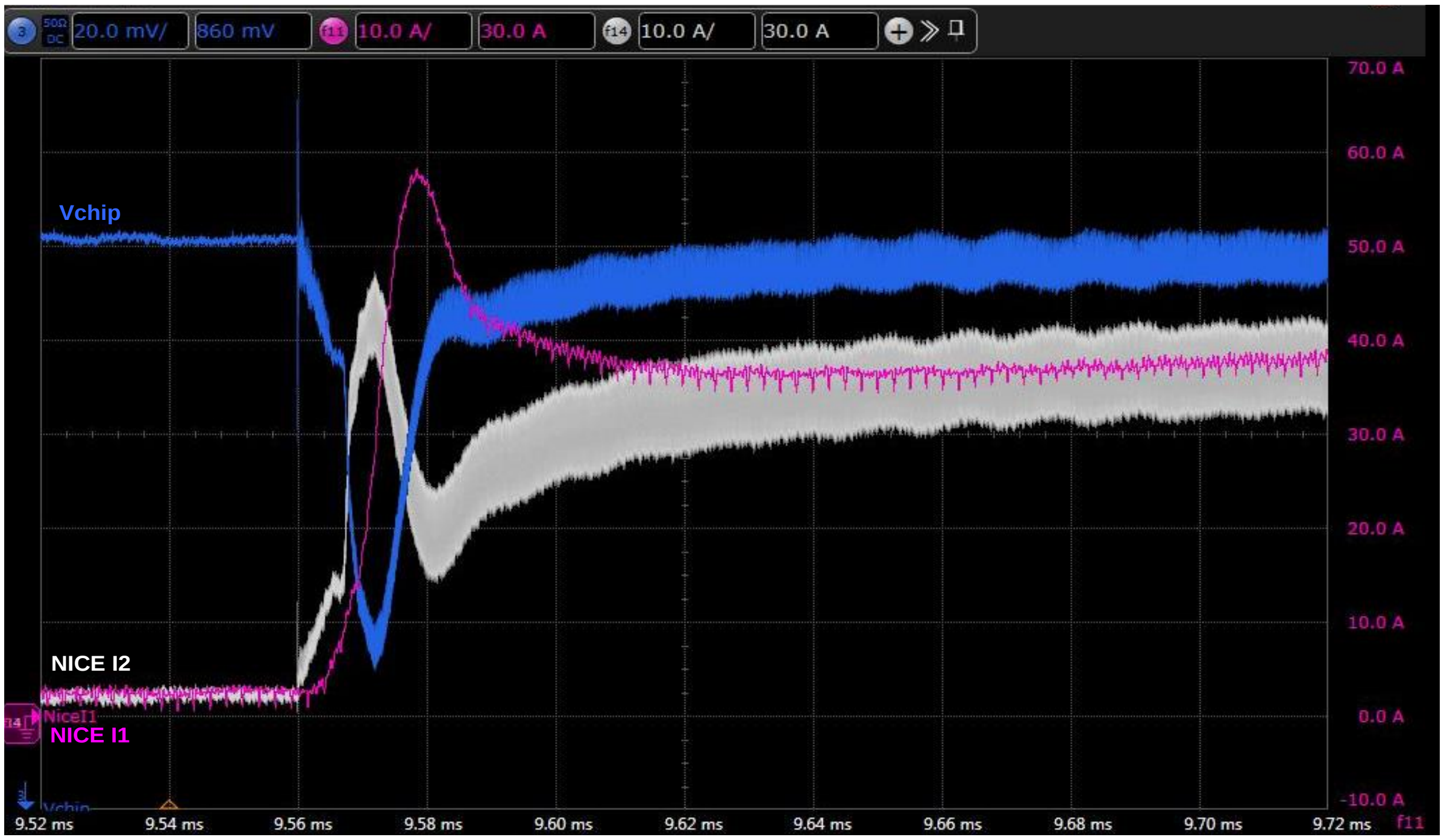

NOV_08 / 2017 / JLF / 46062
MAYO CLINIC SPPDG

**Figure 28 - NICE Estimates for PDN Input Current and FPGA Load Current**

# NICE Limitations

## Component Models

The NICE method relies on the accuracy of the admittance parameters used in the current calculation. In the course of the NICE development and validation, significant issues were discovered regarding the accuracy of ceramic capacitor models when the components are to be used in an operating environment that includes DC-bias and AC-bias, which essentially includes all PDN decoupling applications. We observed that careful validation of board and component models was rewarded with accurate current-waveform results.

## Ground Voltage Gradients

The NICE method relies on a pair of voltage measurements made at separate locations on the DUT board. The measurements for the PDNDUT were carried out using N7020A probes that connect the input SMA-connector ground directly to the oscilloscope ground. When two N7020A probes are connected to the DUT, the “ground” at one sense point is connected through the cable, probe, oscilloscope, second probe, and second cable to the “ground” at the second sense point, forming a “ground-loop.” If large currents flow in the DUT ground conductors between the sense points, then a voltage difference will be generated between the two ground points, and this voltage difference will cause current to flow through the low-impedance path formed by the cables, probes and oscilloscope. This current induces voltage drops along the cable shields that invalidate the measurements.

Ordinary differential probes could be used instead of the N7020A Power Rail probes described herein, and the high common-mode input impedance of a differential probe would mitigate the ground-loop issue. However, this remedy would result in reduced resolution and limited offset range compared to the N7020A.

An analogous problem with ground loops is described in [17] regarding impedance measurements using a VNA. Since the ground-loop current flows on the outside of the coax-cable shield, the associated magnetic field from this current is accessible. By attaching snap-on ferrite EMI suppressors around the coax, the inductance of the shield is increased and the increased shield impedance helps control the magnitude of the ground current.

This remedy works for network analyzers because the output signal from the VNA is an alternating current with an average value of zero. In the course of this effort, it was discovered that ferrite cores along the coax cable were not a good solution when DC currents are involved. Ferrites did increase the shield inductance, and this added inductance behaved as one would expect: it prevented instantaneous changes in the current flowing on the shield. A large DC current was then detected on the shield; thus the problem of shield-induced voltages was exacerbated instead of alleviated.

Any technique or device that limits the current flowing in the ground-loop would help mitigate this error. We note that the J2113A Differential Amplifier from Picotest appears to be a good candidate for this task. When connected between the DUT and the N7020A, the relatively high impedance of the J2113A (>500 ohms between input-ground and output-ground) should effectively break the low-impedance ground-loop path and prevent significant shield-induced voltages from entering the measurement. In this application, the variable input impedance of the N7020A requires special attention to avoid introduction of gain errors at low frequency. We intend to explore this approach.

## Summary

This paper introduced and described a method to calculate time-domain current waveforms from the simultaneous measurement of two time-domain voltage waveforms. Called Non-Invasive Current Estimation (NICE), this method relies on established two-port network theory, together with validated component and board modeling techniques.

NICE was first applied to a simple PDN to demonstrate representative capacitor and board model creation and validation. A frequency-sweep simulation was used to generate tabulated admittance parameters for the combined board assembly and instrumentation, and the resulting Touchstone ® (y2p) file was employed to create a pole-zero approximation for the PDN using code that performs a rational-fitting process. This rational-fit file was saved on the oscilloscope and accessed by the NICE user-defined function. The application of arbitrary load-current profiles resulted in the capture of PDN voltage waveforms by the oscilloscope, and subsequent NICE-function processing displayed a time-correlated current waveform on the oscilloscope. Point-by-point comparison of NICE waveforms to independent current measurements by sense resistor showed an RMS error of less than five percent. This excellent correlation between calculated and measured current waveforms validates the NICE method and code implementation.

NICE was then used to calculate current waveforms for a complex PDN supplying core-logic power to a large multi-pin FPGA. While validation of the NICE method was not demonstrated, the resulting NICE current waveforms were consistent with expectations this PDN and FPGA logic-configuration.

Limitations of the NICE method have been discussed. NICE relies heavily on careful modeling and validation techniques. It was noted that ceramic capacitors can be especially problematic because of the discrepancy between manufacturing-test and application operating conditions. Potential "ground-loop" issues arising from voltage gradients in the DUT return-path conductors may require mitigation.

We demonstrated that with careful characterization and modeling, it was possible to calculate accurate current waveforms without invading the circuit or segregating the supply conductors. Current transients that were previously difficult or impossible to characterize can now be displayed using NICE.

## References

[1] Halliday, David & Resnick, Robert, "Physics Parts I and II", © 1960 by John Wiley & Sons, Inc., p.770
[2] Potter, James & Fich, Sylvan, "Theory of Networks and Lines", © 1963 by Prentice-Hall Inc., p. 11
[3] Frickey, Dean A., "Conversions Between S, Z, Y, h, ABCD, and T Parameters which are Valid for Complex Source and Load Impedances", IEEE Transactions on Microwave Theory and Techniques, Vol. 42, No. 2, February 1994, © 1994 by IEEE
[4] K-Sim online component model generator, [Online], Available: http://ksim.kemet.com/
[5] Murata SimSurfing Design Tool, [Online]. Available: http://www.murata.com/en-us/tool?intcid5=com_xxx_xxx_cmn_nv_xxx
[6] Selection Assistant of TDK Components, [Online], Available: https://product.tdk.com/info/en/technicalsupport/seat/index.html
[7] Taiyo Yuden Design Support Tools, [Online], Available: http://www.yuden.co.jp/or/product/support/index.html#spice

[8] Novak, Istvan, et al, “DC and AC Bias Dependence of Capacitors Including Temperature Dependence”, DesignCon East 2011, [Online], Available: http://www.electrical-integrity.com/Paper_download_files/DCE11_200.pdf
[9] Daniel, Erik S. et al, “Network Analyzer Measurement De-embedding Utilizing a Distributed Transmission Matrix Bisection of a Single THRU Structure”, 63rd ARFTG Conference Digest Spring 2004, p.61-68, © 2004 by IEEE, [Online], Available: http://ieeexplore.ieee.org/document/1387856/
[10] “De-embedding Techniques in Advanced Design System”, © 2008 by Agilent Technologies Inc., [Online], Available: http://literature.cdn.keysight.com/litweb/pdf/5989-9451EN.pdf
[11] “De-embedding and Embedding S-Parameter Networks Using A Vector Network Analyzer, Application Note 1364-1”, © 2004 by Agilent Technologies Inc., [Online], Available: http://anlage.umd.edu/Microwave%20Measurements%20for%20Personal%20Web%20Site/5980-2784EN.pdf
[12] “Simulated Removal & Insertion of Fixtures, Matching and Other Networks”, © 2002 by Anritsu Corp., [Online], Available: https://dl.cdn-anritsu.com/en-us/test-measurement/files/Application-Notes/Application-Note/11410-00278B.pdf
[13] Sandler, Steven, How to Design for Power Integrity: Measuring, Modeling, Simulating Capacitors and Inductors, YouTube video [Online], Available: https://www.youtube.com/watch?v=N4K3y4I4sKA&feature=youtu.be
[14] “S-parameter Simulation and Optimization”, © 2009 by Agilent Technologies Inc., [Online], Available: https://www.utdallas.edu/~rmh072000/Site/Software_and_Links_files/5A_slides.pdf
[15] “Performing Nominal Optimization”, ADS Support Documentation, [Online], Available: http://edadocs.software.keysight.com/display/ads2009/Performing+Nominal+Optimization
[16] “Keysight N7020A Power Rail Probe User’s Guide Fourth Edition”, © 2017 Keysight Technologies Inc., [Online], Available: http://literature.cdn.keysight.com/litweb/pdf/N7020-97003.pdf?id=2535025
[17] Novak, Istvan, ”Measuring MilliOhms and PicoHenrys in Power-Distribution Networks”, DesignCon2000, [Online], Available: http://www.electrical-integrity.com/Paper_download_files/DC00_MeasuringMiliohms.pdf
[18] B. Gustavsen and A. Semlyen, “Rational approximation of frequency domain responses by vector fitting,” Power Delivery, IEEE Transactions on, vol. 14, pp. 1052–1061, 1999. [Online], Available: http://ieeexplore.ieee.org/stamp/stamp.jsp?tp=&arnumber=772353
[19] B. Gustavsen and A. Semlyen, “Enforcing passivity for admittance matrices approximated by rational functions”, IEEE Transactions on Power Systems, vol. 16, no. 1, pp. 97–104, Feb 2001, [Online], Available: http://ieeexplore.ieee.org/stamp/stamp.jsp?tp=&arnumber=910786
[20] B. Gustavsen, "Computer code for rational approximation of frequency dependent admittance matrices", IEEE Trans. Power Delivery, vol. 17, no. 4, pp. 1093-1098, Oct. 2002, [Online], Available: http://ieeexplore.ieee.org/stamp/stamp.jsp?tp=&arnumber=1046889
[21] Pupalaikis, Peter, et al, “Current Sharing Measurements in Multi-Phase Switch Mode DC-DC Converters”, EDICon 2017, [Online], Available: http://www.electrical-integrity.com/Paper_download_files/EDICon2017-PAPER_current-sharing.pdf